\documentclass[prd,nofootinbib,showpacs,preprint]{revtex4-1}
\usepackage[utf8]{inputenc}
\usepackage[T1]{fontenc} % Finnish and Hungarian letters
\usepackage{amsmath}
\usepackage{amsfonts}
\usepackage{amssymb}
\usepackage{amsthm}
\usepackage{booktabs}
\usepackage{cancel}
\usepackage{color}
\usepackage{float}
\usepackage{empheq}
\usepackage{graphicx}
\usepackage{hyperref}
\usepackage{titlesec}
\usepackage[english]{babel}
\usepackage[normalem]{ulem}
\titleformat*{\section}{\Large\bfseries}

\newcommand\bom[1]     {{\mbox{\boldmath $#1$}}}

\newcommand{\lt}{\mathcal{L}} % Lagrangen tiheys
\newcommand{\be}{\begin{equation}}
    \newcommand{\ee}{\end{equation}}
\def\bsp#1\esp{\begin{split}#1\end{split}}
\def\bal#1\eal{\begin{align}#1\end{align}}
\newcommand{\half}{\frac{1}{2}}
\newcommand{\hc}{\text{h.c.}}
\newcommand{\ri}{\text{i}}
\newcommand{\rc}{\text{c}}
\newcommand{\re}{\text{e}}
\newcommand{\rD}{\text{D}}
\newcommand{\rL}{\text{L}}
\newcommand{\rR}{\text{R}}
\newcommand{\rO}{\text{O}}
\newcommand{\tZ}{\theta_Z}
\newcommand{\Ms}{M_{s}}

\newcommand{\MZp}{M_{Z'}}

\newcommand{\da}{\dagger}
\newcommand{\tri}[3]{\left( \begin{array}{c}
        #1 \\ #2 \\ #3
    \end{array}\right)}
\let\oldref\ref
\renewcommand{\ref}[1]{(\oldref{#1})} % Equation references inside parenthesis
\numberwithin{equation}{section}

\begin{document}
    
    \title{Experimental constraints on the neutrino and gauge parameters of the super-weak U(1) extension of the standard model}
    \author{Timo J. Kärkkäinen}
    \email{timo.karkkainen@ttk.elte.hu}
    \affiliation{Institute for Theoretical Physics, ELTE E\"otv\"os Lor\'and University,
        P\'azm\'any P\'eter s\'et\'any 1/A, 1117 Budapest, Hungary}
    \author{Zoltán Trócsányi}
    \email{Zoltan.Trocsanyi@cern.ch}
    \affiliation{Institute for Theoretical Physics, ELTE E\"otv\"os Lor\'and University,
        P\'azm\'any P\'eter s\'et\'any 1/A, 1117 Budapest, Hungary and \\
        ELKH-DE Particle Physics Research Group, University of Debrecen \\ 
        4010 Debrecen, PO Box 105, Hungary}
    \date{26.05.2021}
    
    \begin{abstract}
        The super-weak force is a minimal, anomaly-free U(1) extension of 
        the standard model (SM), designed to explain the origin of 
        (i) neutrino masses and mixing matrix elements, (ii) dark matter, 
        (iii) cosmic inflation, (iv) stabilization of the electroweak vacuum  
        and (v) leptogenesis. 
        We discuss the neutrino sector of this model in detail and study 
        the allowed parameter space of the neutrino Yukawa matrices 
        and mixing matrix elements. The model generates nonstandard 
        neutrino interactions, whose allowed experimental limits are
        used to constrain the parameter space of the model. We provide 
        benchmark points in the relevant parameter space that fall within 
        the sensitivity region of the SHiP and MATHUSLA experiments.
    \end{abstract}
    
    \maketitle 
    \tableofcontents
    \clearpage
    \pagebreak
    
    \section{Introduction}
    
    Finding all the building blocks of the standard model (SM) of 
    elementary particle interactions has been an effort spanning 
    nearly half a century, concluding with the discovery of the 
    Higgs boson in 2012 \cite{Aad:2012tfa,Chatrchyan:2012xdj}.
    During that time several hints and discoveries have unfolded, 
    confirming that the SM cannot describe all observations in 
    particle physics. In this post-SM era, several simple extensions 
    of the SM are experiencing a renaissance as the number of reported
    deviations from the SM in particle physics are increasing \cite{Aguilar-Arevalo:2018gpe,Abi:2021gix}. 
    The discovery of the in-flight flavour oscillation of neutrinos
    \cite{Fukuda:1998mi,Ahmad:2001an} in vacuum and matter requires
    that at least two of the three known neutrinos must have 
    non-vanishing masses. However, the SM is devoid of a suitable 
    mass generation mechanism for the neutrinos. To counter 
    this, the existence of right-handed neutrinos that are sterile under the standard 
    model gauge group have been proposed. The seesaw mechanism 
    \cite{Yanagida:1979as,GellMann:1980vs,Mohapatra:1979ia,Schechter:1980gr,Glashow:1979nm}
    utilizes these neutrinos that mix weakly with the active ones.
    Sterile neutrinos have been considered as dark matter candidates
    \cite{Asaka:2005pn,Shaposhnikov:2006xi,Canetti:2012kh,Boyarsky:2018tvu,Boyarsky:2018tvu,Adhikari:2016bei,Abazajian:2001nj,Abazajian:2012ys,Dolgov:2000ew,Akhmedov:1998qx,Iwamoto:2021fup}, 
    and their role for the baryonic asymmetry of the universe via 
    leptogenesis has been studied intensively, see for instance  \cite{Asaka:2005pn,Canetti:2012kh,Fukugita:1986hr}
    %Pilaftsis:2003gt,Pilaftsis:2005rv,Davidson:2008bu,Blanchet:2009kk}. 
    An unidentified 3.5 keV X-ray line discovered in Andromeda galaxy 
    and several galaxy clusters \cite{Bulbul:2014sua,Boyarsky:2014jta} 
    may be interpreted to originate from sterile neutrino decays
    at one-loop order in perturbation theory, with the mass of sterile 
    neutrino being about $7.1$ keV\footnote{We use natural units in this paper.}
    and mixing very weakly with the active ones, 
    with mixing angle $\rO(10^{-11})$.
    
    Light sterile neutrinos at keV range have been searched by several experiments
    \cite{Atre:2009rg} measuring the electron energy spectrum of unstable $\beta$ 
    decaying nuclei. The sterile component of $\nu_\re$ would cause missing events 
    and a kink concentrated at $E^e_{\max} - m_{\rm s}$, where $m_{\rm s}$ is the mass of sterile 
    neutrino and $E^e_{\max}$ is the endpoint of electron energy spectrum. These 
    \textit{kink searches} have ruled out a mixing larger than 
    $|U_{e4}|^2 = \rO(10^{-3})$. MeV scale neutrinos are sensitive to peak searches, where a meson decay is enhanced by the sterile neutrino. Mixing larger than $|U_{e4}|^2 = \rO(10^{-5})$ has been ruled out on 10--100\,MeV mass range \cite{Britton:1992pg,Britton:1992xv,Yamazaki:1984sj,Artamonov:2014urb}.
    Even heavier neutrinos of mass $\rO(1)$\,GeV 
    are constrained by beam dump experiments, with CHARM constraining 
    $|U_{e4}|^2 \lesssim 10^{-6}$ for sterile neutrinos in the 0.5--2\,GeV mass range 
    \cite{Bergsma:1985is}. Searches for sterile neutrinos of larger masses in
    decays of the $Z^0$ boson have been performed by DELPHI 
    \cite{Abreu:1996pa}, constraining $|U_{e4}|^2 \lesssim 10^{-5}$ for sterile 
    neutrino mass range 2--60\,GeV.
    
    New physics in the neutrino sector can be parameterized via the 
    nonstandard interaction (NSI) formalism \cite{Grossman:1995wx}, 
    which produces distortions to neutrino oscillation spectra and 
    neutrino scattering experiments. The NSI operators can be generated 
    at the effective theory level in theories which couple new fields 
    to active neutrinos. Particularly a new U(1) gauge boson may couple 
    to neutrinos if they have suitable U(1) charge assignments.
    In this paper we explore such a possibility within the
    model of the super-weak force \cite{Trocsanyi:2018bkm}. We study how
    the experimental limits on the NSI constrain the parameter space.
    We also demonstrate that the model provides benchmark points in the 
    allowed parameter region that fall in the sensitivity region of future 
    experiments searching for sterile neutrinos.
    
    The structure of this paper is as follows. In Sec. \ref{Sec2}, 
    we present a description of the super-weak model. Neutrino mass 
    generation, mixing and sub-leading corrections to neutrino mass are
    covered in Sec. \ref{Sec3}. Nonstandard 
    neutrino interactions are derived and constrained in Sec. \ref{Sec5}. 
    We present our benchmark points and results in Sec. \ref{Sec6}, 
    and give our conclusions in Sec. \ref{Sec7}.
    
    \section{Model description\label{Sec2}}
    
    In the super-weak extension of the SM \cite{Trocsanyi:2018bkm}, the
    field content is enlarged by one complex singlet scalar and three
    right-handed fermion fields that become massive sterile neutrinos after
    spontaneous symmetry breaking. Here we recall the details of the gauge,
    scalar and fermion sectors to the extent we need for our analysis.
    
    \subsection{Gauge extension}
    The underlying gauge group of the model is
    \be 
    G =
    \text{SU}(3)_\rc \otimes \text{SU}(2)_\rL \otimes \text{U}(1)_Y \otimes \text{U}(1)_z
    \,,
    \ee 
    which is anomaly-free with the quantum numbers given in
    Table~\ref{tab:Field-rep}.
    The covariant derivative for the matter field $f$ is defined as
    \be 
    D_f^\mu = \partial^\mu 
    + \ri g_\rL \bom{T}\cdot \textbf{W}^\mu 
    + \ri y_f g_y B^\mu + \ri z_f g_z Z^\mu,
    \ee 
    where $g_\rL$, $g_y$ and $g_z$ are the gauge couplings of 
    corresponding gauge groups. (We omit the SU(3)$_\rc$ part in our 
    description, since it is not relevant in our discussion.) 
    The corresponding gauge fields are
    $\textbf{W}^\mu = (W_1^\mu, W_2^\mu, W^3_\mu)$, $B^\mu$ 
    and $Z^\mu$. The charges $y_f$ and $z_f$ correspond to U(1)$_Y$ 
    and U(1)$_z$ groups, and $\bom{T} = (T^1,T^2,T^3)$ are the
    generators of the SU(2)$_\rL$ group (1/2 times the Pauli matrices). 
    \begin{table}[th] % \rule gives extra margins for \frac(tions)
        \caption{\label{tab:Field-rep}Gauge group representations
            and charges of the fermions and scalars in the super-weak model}
        \begin{tabular}{|c||cccc|}\hline 
            \textbf{Field} &     SU(3)$_\rc$ & SU(2)$_\rL$ & U(1)$_Y$ & U(1)$_z$ \\ \hline \hline 
            \rule{0pt}{3ex}$Q_\rL$      &  \textbf{3} & \textbf{2} & $\frac{1}{6}$& $\frac{1}{6}$\\\hline 
            \rule{0pt}{3ex}$u_\rR$       & \textbf{3} & \textbf{1} & $\frac{2}{3}$& $\frac{7}{6}$\\\hline 
            \rule{0pt}{3ex}$d_\rR$       & \textbf{3} & \textbf{1} & $-\frac{1}{3}$& $-\frac{5}{6}$\\\hline 
            \rule{0pt}{3ex}$L_\rL$       & \textbf{1} & \textbf{2} & $-\half$& $-\frac{1}{2}$\\\hline 
            \rule{0pt}{3ex}$\ell_\rR$    & \textbf{1} & \textbf{1} & $-1$& $-\frac{3}{2}$\\\hline 
            \rule{0pt}{3ex}$N_\rR$       & \textbf{1} & \textbf{1} & $0$& $\frac{1}{2}$\\\hline 
            \rule{0pt}{3ex}$\phi$      & \textbf{1} & \textbf{2} & $\half$& 1\\\hline 
            \rule{0pt}{3ex}$\chi$      & \textbf{1} & \textbf{2} & $0$&  $-1$\\\hline 
        \end{tabular}
    \end{table}
    
    The gauge kinetic terms of the Lagrangian,
    \be 
    \lt_\text{gauge} = -\frac{1}{4}B^{\mu\nu}B_{\mu\nu} - \frac{1}{4}Z^{\mu\nu}Z_{\mu\nu} - \frac{1}{4}\textbf{W}^{\mu\nu}\cdot \textbf{W}_{\mu\nu} - \frac{\varepsilon}{2}B_{\mu\nu}Z^{\mu\nu}
    \ee 
    include a kinetic mixing term, proportional to a real 
    parameter $\varepsilon \ll 1$. We can redefine the U(1) fields 
    via a linear transformation
    \be 
    \binom{\hat{B}_\mu}{\hat{Z}_\mu} = \left( \begin{array}{cc}
        1 & \sin \theta_\varepsilon \\ 0 & \cos \theta_\varepsilon
    \end{array}\right) \binom{B_\mu}{Z_\mu}, \quad \sin \theta_\varepsilon \equiv \varepsilon
    \,.
    \ee 
    Then the covariant derivative can be rewritten as
    \be 
    D_f^\mu = \partial^\mu 
    + \ri g_\rL \bom{T}\cdot \textbf{W}^\mu 
    + \ri (y, z)_f
    \left( \begin{array}{cc} g_y & -g_y' \\ 0 & g_z'\end{array}\right) 
    \binom{\hat{B}_\mu}{\hat{Z}_\mu},
    \ee 
    where $g_y' = g_y\tan \tZ$ and $g_z' = g_z/\cos \tZ$. In this
    basis the kinetic mixing is absent, which can be achieved but at an
    arbitrarily chosen scale. The scale dependence of the mixing term is
    mild \cite{Iwamoto:2021fup}, hence we neglect it in this exploratory
    paper and will study its effect in a more complete analysis. The $\hat Z$
    and $\hat B$ fields  being both electrically neutral can  still mix
    with the neutral $W_3$ boson via a rotation:
    \be 
    \tri{W_\mu^3}{\hat{B}_\mu}{\hat{Z}_\mu} = \begin{pmatrix}
        \cos \theta_W \cos \tZ & \cos \theta_W \sin \tZ & \sin \theta_W\\
        -\sin \theta_W \cos\tZ & -\sin \theta_W \sin \tZ & \cos \theta_W\\
        -\sin \tZ & \cos \tZ & 0
    \end{pmatrix}\tri{Z_\mu}{Z'_\mu}{A_\mu},
    \ee 
    where the Weinberg angle $\theta_W$ is defined as in the SM.  The
    second mixing angle $\tZ$ mixes the massive fields and is
    determined by the relation
    \be 
    \bsp
    \tan 2\tZ &=
    -\frac{2(2g_z'-g_y')\sqrt{g_\rL^2+g_y^2}}
    {g_\rL^2+g_y^2-(2g_z^{\prime}-g_y')^2 - 4g_z^{\prime 2}\tan^2 \beta}
    %\\&
    %-\frac{2(2g_z'-g_y')\cos\theta_W}
    %{1-(2g_z^{\prime}-g_y')^2 \cos^2\theta_W - 4g_z^{\prime 2} \cos^2\theta_W\tan^2 \beta}
    \esp
    \ee 
    where $\tan\beta = w/v$ is the ratio of the vacuum expectation values
    (VEVs) of the scalar fields (see Eq.~(\ref{vevs})).  The extra degree
    of freedom introduced by the U(1)$_z$ group manifests itself as an
    extra neutral massive gauge boson, which we call the $Z'$ boson. 
    It mixes with the $Z$ boson, the other massive neutral gauge boson in
    the model. 
    
    \subsection{New neutral currents}
    
    All fermions have $z$-charges, which lead to new neutral currents
    $\bar{f}\,\Gamma_{Z'ff}^\mu f$ coupled to the $Z'_\mu$.  The coupling of 
    the $Z'$ boson to fermion fields $f = \nu$, $\ell$, $u$, $d$ has the form
    \cite{Trocsanyi:2018bkm}
    \be
    \label{T-coupling}
    \Gamma_{Z'ff}^\mu =  -\ri e\gamma^\mu  (C_{Z'ff}^\rR P_\rR + C_{Z'ff}^\rL P_\rL)
    \ee
    where $P_{\rL,\rR} = \frac12(1\mp\gamma_5)$ are chiral projection operators and
    \be
    C_{Z'ff}^\rR = -g_f^+\sin \tZ + h_f^+\cos \tZ
    \,,\quad
    C_{Z'ff}^\rL = -g_f^- \sin \tZ + h_f^- \cos \tZ
    \,.
    \ee
    The coupling factors are
    \be \label{T-coeff}
    g_f^+ = -\frac{\sin \theta_W}{\cos \theta_W} e_f
    \\,\quad 
    g_f^- = \frac{T_f^3 - \sin^2\theta_W e_f}{\sin \theta_W \cos \theta_W}
    \,, \quad 
    h_f^\pm = \frac{g'_Z R_f^\pm
        + (g_z'-g_y')(e_f - R_f^\mp)}{g_\rL \sin \theta_W},
    \ee 
    with values of $e_f, T^3_f, R_f^+$ and $R_f^-$ given in
    Table~\ref{coupling-table}. Explicit computations yield
    \begin{align}
        eC_{Z'\nu\nu}^\rL &= 
        \frac{1}{2} (g_y'-g_z') \cos \tZ
        - \frac{g_\rL}{2} \frac{\sin \tZ}{\cos \theta_W}
        \,,\qquad
        eC_{Z'\nu\nu}^\rR =\frac{g_z'}{2} \cos \tZ
        \,,\\
        eC_{Z'\ell\ell}^\rL &=
        \frac{1}{2} (g_y'-g_z') \cos \tZ
        + \frac{g_\rL}{2} \frac{1-2\sin^2 \theta_W}{\cos \theta_W} \sin \tZ
        \,,\\
        eC_{Z'\ell\ell}^\rR &= 
        \left(g_y'-\frac{3}{2} g_z'\right) \cos \tZ
        -g_\rL \frac{\sin^2 \theta_W}{\cos \theta_W} \sin \tZ
        \,,\\
        eC_{Z'{uu}}^\rL &= 
        -\frac{1}{6} (g_y'-g_z') \cos \tZ
        -\frac{1}{6} g_\rL \frac{3-4\sin^2 \theta_W}{\cos \theta_W} \sin \tZ
        \,,\\
        eC_{Z'{uu}}^\rR &= 
        -\frac{1}{6} (4g_y'-7g_z') \cos \tZ
        +\frac{2}{3} g_\rL \frac{\sin^2 \theta_W}{\cos \theta_W} \sin \tZ
        \,,\\
        eC_{Z'{dd}}^\rL &= 
        -\frac{1}{6} (g_y'-g_z') \cos \tZ
        +\frac{1}{6} g_\rL \frac{3-2\sin^2 \theta_W}{\cos \theta_W} \sin \tZ
        \,,\\
        eC_{Z'{dd}}^\rR &= 
        \frac{1}{6} (2g_y'-5g_z') \cos \tZ
        -\frac{1}{3} g_\rL\frac{\sin^2 \theta_W}{\cos \theta_W} \sin \tZ
    \end{align}
    The couplings $C_{Z'ff}^{\rL/\rR}$ are real because the gauge couplings are
    themselves real.
    \begin{table}[t]
        \centering 
        \caption{\label{coupling-table}Table of all the couplings for different fields.}
        \begin{tabular}{|c||c|c|c|c|}
            \hline 
            \hline 
            \textbf{field}  & $e_f$ & $T_f^3$ & $R_f^+$ & $R_f^-$ \\\hline \hline 
            $u,\:c,\:t$     & $\frac{2}{3}$ & $\half$ &$ \half$ & 0\\\hline 
            $d,\:s,\:b$     & $-\frac{1}{3}$ & $-\half$ & $-\half$ & 0\\\hline 
            $\nu_\re, \nu_\mu,\nu_\tau$  & $0$ & $\half$ & $\half$ &0 \\\hline 
            e$^-,\mu^-,\tau^-$& $-1$ & $-\half$& $-\half$ & 0\\\hline 
            \hline
        \end{tabular}
    \end{table}
    
    \subsection{Scalar extension}
    
    The scalar sector of the model consists of a complex doublet 
    $\phi = \binom{\phi^+}{\phi^0}$ and a complex singlet $\chi$. 
    The scalar Lagrangian can be written as
    \be 
    \lt_\text{scalar} = (D_\mu \phi)^*(D^\mu \phi) + |D_\mu \chi|^2 
    - \mu_\phi^2\phi^\da \phi - \mu_\chi^2|\chi|^2 
    - \lambda_\phi (\phi^\da \phi)^2 - \lambda_\chi |\chi|^4 - \lambda (\phi^\da \phi) |\chi|^2,
    \ee 
    up to a constant term irrelevant to our discussion. All the couplings 
    are real. The portal coupling $\lambda$ induces mixing between
    the neutral component of $\phi$ and $\chi$. The potential is 
    minimal at the scalar field values
    \be 
    \phi_0 = \frac{1}{\sqrt{2}}\binom{0}{v}
    \,, \quad 
    \chi_0 = \frac{w}{\sqrt{2}}
    \,,
    \ee 
    where
    \be \label{vevs}
    \frac{v}{\sqrt{2}} =
    \sqrt{\frac{2\lambda_\chi \mu_\phi^2- \lambda \mu_\chi^2}
        {4\lambda_\phi \lambda_\chi - \lambda^2}}
    \,,\quad 
    \frac{w}{\sqrt{2}} =
    \sqrt{\frac{2\lambda_\phi \mu_\chi^2- \lambda \mu_\phi^2}
        {4\lambda_\phi \lambda_\chi - \lambda^2}}
    \,.
    \ee 
    In this paper we perform our analyses at tree level, hence we can
    choose the unitary gauge,  in which after spontaneous symmetry
    breaking (SSB) the fields $\phi$ and $\chi$ can be parametrized in
    terms of two real scalar fields $h'$ and $s'$ as
    \be 
    \phi = \frac{1}{\sqrt{2}}\binom{0}{v+h'}
    \text{~~and~~}
    \chi = \frac{1}{\sqrt{2}}(w+s')
    \,.
    \ee 
    The mass matrix for scalars in $(h',s')$ basis is given by
    \be
    M^\text{scalar}_{ij} =
    \left( \begin{array}{cc}
        3\lambda_\phi v^2 + \half \lambda w^2 + \mu_\phi^2 & \lambda vw 
        \\ 
        \lambda vw & 3\lambda_\chi w^2 + \half \lambda v^2 + \mu_\chi^2
    \end{array}\right),
    \ee
    which has eigenvalues
    \be 
    M_{h_\pm}^2 = \lambda_\phi v^2 + \lambda_\chi w^2 \pm \sqrt{(\lambda_\phi v^2 - \lambda_\chi w^2)^2 + (\lambda vw)^2}.
    \ee 
    We adopt the convention $M_{h_-} \leq M_{h_+}$. Diagonalizing the mass
    matrix with an orthogonal rotation $O_S$, we obtain the mass eigenstates
    $h$ and $s$:
    \be 
    \binom{h}{s} = \left( \begin{array}{cc}
        \cos \theta_S & -\sin \theta_S \\ \sin \theta_S & \cos \theta_S 
    \end{array}\right) \binom{h'}{s'}
    \ee
    where the scalar mixing angle is given by
    \be 
    \tan 2\theta_S = -\frac{\lambda vw}{\lambda_\phi v^2 - \lambda_\chi w^2}
    \equiv
    -\frac{\lambda \tan\beta}{\lambda_\phi - \lambda_\chi\tan^2\beta}
    \,,
    \ee 
    with $\tan\beta=w/v$. The correspondence between the states
    $(h_+,h_-)$ and $(h,s)$ depends on the sign of $\theta_S$, which can be
    determined experimentally by the hierarchy between the masses of the Higgs
    particle and the new scalar particle. It was shown in Ref.~\cite{Peli:2019vtp}
    that the stability of the vacuum is assured up to the Planck scale 
    within the super-weak model if the SM Higgs particle is lighter than the 
    new scalar.
    
    \subsection{Neutrino extension}
    
    The fermion sector of the SM is extended by three massive right-handed
    neutrinos that are charged under U(1)$_z$
    only, i.e.~sterile under the SM gauge group. They can have either Dirac
    or Majorana nature. Since the Yukawa term that produces the Majorana
    mass term for Dirac-type neutrinos is allowed by the gauge symmetries,
    we include both Dirac- and Majorana-type Yukawa terms in the
    Lagrangian.
    
    We write the leptonic part of the Yukawa Lagrangian as
    \be
    -\lt_Y^\ell = \half \overline{(N_\rR)^{c}}\:Y_N \chi^*\:N_\rR
    + \overline{L_\rL}\:\phi Y_\ell\:\ell_\rR
    + \overline{L_\rL}\:\phi^c Y_\nu\:N_\rR 
    + \hc
    \ee
    where the fermion fields are triplets in flavour space,
    \be
    N_\rR = \tri{N_1}{N_2}{N_3}
    \,,\quad
    L_{\rL} = \tri{L_{\rL\re}}{L_{\rL\mu}}{L_{\rL\tau}}
    \,,\quad
    \ell_\rR = \tri{e_\rR}{\mu_\rR}{\tau_\rR}
    \,.
    \label{eq:Yukawa}
    \ee
    The superscript $c$ denotes charge conjugation, $\varphi^c = \ri\sigma_2
    \varphi^*$ for a generic two-component field $\varphi$ (left-handed Weyl spinor or
    complex scalar doublet).
    
    We can \emph{choose} to work on a basis where the $Y_N$ and $Y_\ell$ are diagonal
    matrices, corresponding to mass eigenstates of the lepton fields $N$ and $\ell$.
    Before diagonalization, i.e.~on the flavour basis, the Majorana-type
    Yukawa matrix $Y_N'$ was symmetric, so the diagonalizing matrix $O_N$
    is orthogonal.  In contrast, the charged lepton Yukawa matrix
    $Y_\ell'$ is diagonalized via biunitary transformation, using two
    unitary matrices $U_{\ell \rL}$ and $U_{\ell \rR}$:
    \be 
    Y_N = O_{N}^TY_N'O_{N}, \quad Y_\ell' = U_{\ell \rL}^\da Y_\ell' U_{\ell \rR}.
    \ee 
    In this context, we define the corresponding Dirac neutrino Yukawa
    matrix to be $Y_\nu \equiv Y'_\nu O_N$ without losing generality.
    
    After SSB, the Yukawa Lagrangian of the leptons becomes 
    \be
    \label{lepton-yukawa}
    -\lt_Y^\ell = \frac{w+s'}{2\sqrt{2}}
    \overline{(N_\rR)^{c}}\:Y_N\:N_\rR
    + \frac{v+h'}{\sqrt{2}}
    \overline{\ell_\rL}\:Y_\ell\:\ell_\rR
    + \frac{v+h'}{\sqrt{2}}
    \overline{\nu_\rL}\:Y_\nu\:N_\rR 
    + \hc\,.
    \ee

    \section{Neutrino masses and mixing\label{Sec3}}
    
    In this section we review a possible parametrization of the type-I 
    seesaw mechanism within the super-weak model. While it is very similar 
    to the standard in the literature, we present it explicitly in the form that we
    shall use later to find benchmark points. 
    
    \subsection{Seesaw expansion}
    
    First we identify the neutrino mass matrix from Eq.~(\ref{lepton-yukawa}),
    and write it in the full $6\times 6$ form:
    \begin{align}
        -\lt_m^\nu &= \half \binom{\nu_\rL}{(N_\rR)^c}^T\left( \begin{array}{cc}
            0 & m_\rD \\ m_\rD^T & m_\rR
        \end{array}\right) \binom{(\nu_\rL)^c}{N_\rR}+\hc  
        \,,
    \end{align}
    where $\nu_\rL$ is a similar triplet as $N_\rR$ defined in Eq.~\eqref{eq:Yukawa},
    while the Dirac and Majorana neutrino mass terms are
    \be 
    m_\rD = \frac{v}{\sqrt{2}}Y_\nu
    \text{~~and~~}
    m_\rR = \frac{w}{\sqrt{2}}Y_N
    \,.
    \ee 
    Then we can block-diagonalize the block mass matrix up to small corrections:
    % \be
    % \bsp
    % \label{block-diag}
    % \left( \begin{array}{cc}
    %     m_\nu & 0 \\ 0 & m_N
    % \end{array}\right)  &=
    % \left( \begin{array}{cc}
    % I & -m_\rD m_\rR^{-1} \\ m_\rR^{-1}m_\rD^\da & I
    % \end{array}\right) 
    % \left( \begin{array}{cc}
    %     0 & m_\rD \\ m_\rD^T & m_\rR
    % \end{array}\right) 
    % \left( \begin{array}{cc}
    %     I & m_\rD^*m_\rR^{-1\dagger} \\ -m_\rR^{-1}m_{D}^T & I
    % \end{array}\right)
    % \\ &
    % = \left( \begin{array}{cc}
    % -m_\rD m_\rR^{-1}m_\rD^T & -m_\rD m_\rR^{-1}m_\rD^Tm_\rD^*m_\rR^{-1\dag}  \\ -m_\rR^{-1}m_\rD^\da m_\rD m_\rR^{-1}m_\rD^T & m_\rR + m_\rD^Tm_\rD^* m_\rR^{-1\dag} + m_\rR^{-1}m_\rD^\da m_\rD
    % \end{array}\right)
    % \esp
    % \ee
    \be
    \bsp
    \label{block-diag}
    \left(\!\begin{array}{cc}
        m_\nu & 0 \\ 0 & m_N
    \end{array}\!\right)  &=
    %\left( \begin{array}{cc}
    %I & -U_{\rm as}^* \\ U_{\rm as}^T & I
    %\end{array}\right) 
    \left(\!\begin{array}{cc}
        I & U_{\rm as} \\ -U_{\rm as}^\da & I
    \end{array}\!\right)^T
    \left(\!\begin{array}{cc}
        0 & m_\rD \\ m_\rD^T & m_\rR
    \end{array}\!\right) 
    \left(\!\begin{array}{cc}
        I & U_{\rm as} \\ -U_{\rm as}^\da & I
    \end{array}\!\right)
    \approx \left(\!\begin{array}{cc}
        -m_\rD m_\rR^{-1}m_\rD^T & 0 \\ 0 & m_\rR
    \end{array}\!\right) 
    \esp
    \ee 
    where 
    \be 
    U_{\rm as} = m_\rD^*m_\rR^{-1\dag}
    =  \left( \begin{array}{ccc}
        U_{e 4} & U_{e 5} & U_{e 6} \\ U_{\mu 4} & U_{\mu 5} & U_{\mu 6} \\ U_{\tau 4} & U_{\tau 5} & U_{\tau 6}
    \end{array} \right) 
    \,,
    \label{eq:Uas}
    \ee
    so
    \be 
    m_\rD = U_{\rm as}^* m_\rR 
    \,,\qquad
    m_\rD^T = m_\rR U_{\rm as}^\da
    \,.
    \label{eq:Uas}
    \ee
    In the last step of \eqref{block-diag} we neglected blocks suppressed by powers 
    of $m_\rD$, such as the off-diagonal blocks $U_{\rm as}^* m_\rD^T U_{\rm as}$
    %$=-m_\rD m_\rR^{-1}m_\rD^Tm_\rD^*m_\rR^{-1\dag}$ 
    and its transpose. Those matrices have small elements for all benchmark 
    points in our discussion during the next sections essentially because
    there are stringent bounds on the elements of the active-sterile mixing 
    matrix in the mass range 1--80\,GeV of the sterile neutrinos: 
    $|U_{a i}|^2\lesssim 10^{-5}$ $(a = e,\:\mu,\:\tau$, $i = 4,\:5,\:6)$ 
    \cite{Alekhin:2015byh,Bondarenko:2018ptm}. 
    
    We also neglected terms in the seesaw expansion from the diagonal block $m_N$.
    The order of magnitude of such a correction to a sterile neutrino of
    mass in the keV range is
    \begin{align} \label{eq:m4-corr}
        \Delta m_N &\sim \left(m_\rD^Tm_\rD^*m_\rR^{-1}\right)_{1i} 
        %= \frac{v^2}{2} \sum \limits_{j,k=1}^3(Y_\nu^T)_{1k} (Y_\nu^*)_{kj} (m_\rR^{-1})_{ji} 
        = \frac{v^2}{2} \frac{(Y_\nu^T Y_\nu^*)_{1i}}{(m_\rR)_{ii}}
        %\\&
        \simeq \frac{(Y_\nu^T Y_\nu^*)_{1i}}{10^{-19}}  \times  \frac{\text{keV}}{(m_N)_{ii}} \times 6\,\text{eV},
    \end{align} 
    which is much smaller than the keV scale for all our benchmarks with one 
    exception where numerical coincidence leads to moderately large correction 
    of 1.3\,keV. (The actual size of the correction terms can be calculated 
    using our benchmark Yukawa matrices $Y_\nu$, with values given in Sect.~\ref{Sec6}.) 
    The corrections to a GeV sterile neutrino mass are larger by one order of
    magnitude.  Thus we can safely drop the neglected terms.
    
    A similar term emerges also for the light neutrino mass matrix as 
    sub-leading order terms from seesaw expansion. Within second order, 
    these are \cite{Grimus:2000vj}
    \be 
    \Delta m_\nu = \half m_\nu m_\rD^\da m_N^{-2} m_\rD
    + \half \left( m_\nu m_\rD^\da m_N^{-2}m_\rD\right)^T
    \ee 
    By direct calculation for the benchmark points, this correction turns 
    out to be much smaller than $10^{-7}$\,eV  (see Table \ref{BPtable}), 
    so our approximations in the seesaw expansion are justified. As a result 
    $m_\nu \approx -m_\rD m_\rR^{-1}m_\rD^T$ and $m_N \approx m_\rR$ 
    are the approximate mass matrices for active and sterile neutrinos. 
    
    \subsection{Diagonalization of the light neutrino mass matrix at one loop}
    
    At this point $m_N$ is already diagonal, but $m_\nu$ is not so, 
    so next we diagonalize the light neutrino mass matrix $m_\nu$:
    \be 
    \left( \begin{array}{cc}
        U_2 & 0 \\ 0 & I
    \end{array}\right)^T \left( \begin{array}{cc}
        m_\nu & 0 \\ 0 & m_N
    \end{array}\right)  \left( \begin{array}{cc}
        U_2 & 0 \\ 0 & I
    \end{array}\right)  = \left( \begin{array}{cc}
        U_2^T m_\nu U_2 & 0 \\ 0 &  m_N 
    \end{array}\right) = \left( \begin{array}{cc}
        m_\nu^\text{diag} & 0 \\ 0 & m_N^\text{diag} 
    \end{array}\right) ,
    \label{eq:fulldiag}
    \ee 
    where $U_2$ is a $3\times 3$ unitary matrix. We have experimental 
    constraints on the upper limits the elements of $m_\nu^\text{diag}$
    \cite{Aghanim:2018eyx,Aker:2019uuj}. Even if the tree-level matrix
    $m_\nu^\text{diag}$ satisfies those limits, one has to check 
    that the inclusion of loop corrections do not upset them.
    
    Due to the active-sterile mixing of neutrinos, the one-loop
    correction to light neutrino masses has contributions involving
    right-handed neutrinos and neutral scalars or vector bosons in the loop.
    The required computations in the context of gauged U(1) extensions of
    the standard model is documented in Ref.~\cite{Iwamoto:2021wko}. Here we
    simply recall the total one-loop correction for the super-weak model:
    \be
    \bsp
    (\delta m_\nu)_{ij} &=\sum _{V=Z,Z'}
    \frac{3e^2}{16\pi^2M_V^2}\Big(C^\rL_{V\nu\nu}-C^\rR_{V\nu\nu}\Big)^2
    \sum_{n=1}^6(U_\rL^*)_{in}m_n^3
    \frac{\ln \frac{m_n^2}{M_V^2}}{\frac{m_n^2}{M_V^2}-1}(U_\rL^\dagger)_{nj}
    \\ &
    +\frac{1}{16\pi^2v^2} \sum_{n=1}^6(U_\rL^*)_{in}m_n^3(U_\rL^\dagger)_{nj}
    \left(\sin^2\theta_S\frac{\ln \frac{m_n^2}{\Ms^2}}{\frac{m_n^2}{\Ms^2}-1}
    + \cos^2\theta_S\frac{\ln \frac{m_n^2}{M_h^2}}{\frac{m_n^2}{M_h^2}-1}\right)
    \esp
    \ee
    where the summation runs over all six neutrinos%
    \footnote{Here $\nu_i = N_{i-3}$ for $i>3$.} and the
    $U_\rL = (U_2\; U_{\rm as})$ matrix is the upper $(3\times 6)$ part of the
    unitary matrix that diagonalizes the full neutrino mass matrix.
    \begin{figure}[ht!]
        \begin{center}
            \includegraphics[width=0.7\linewidth]{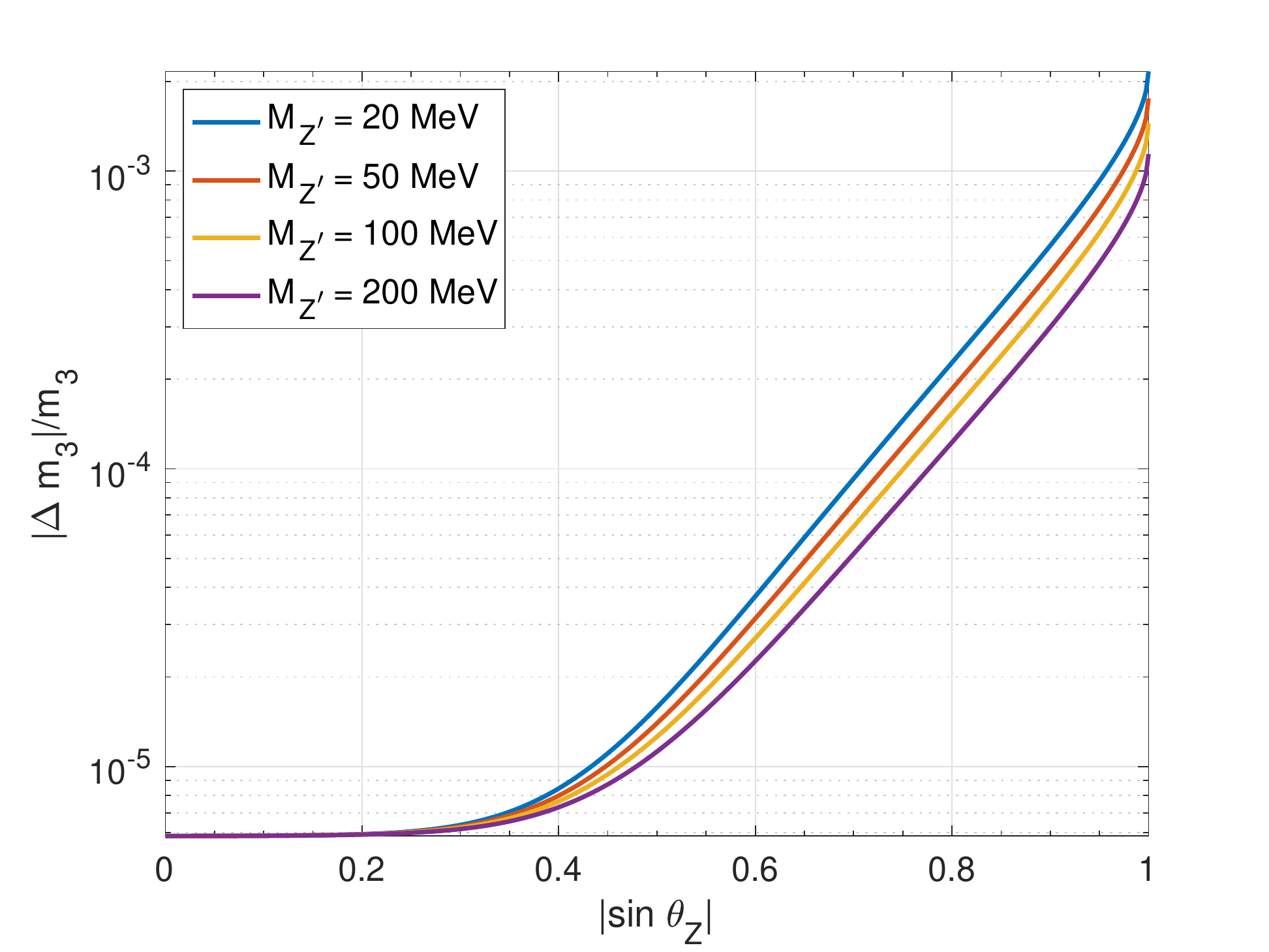}
            \includegraphics[width=0.7\linewidth]{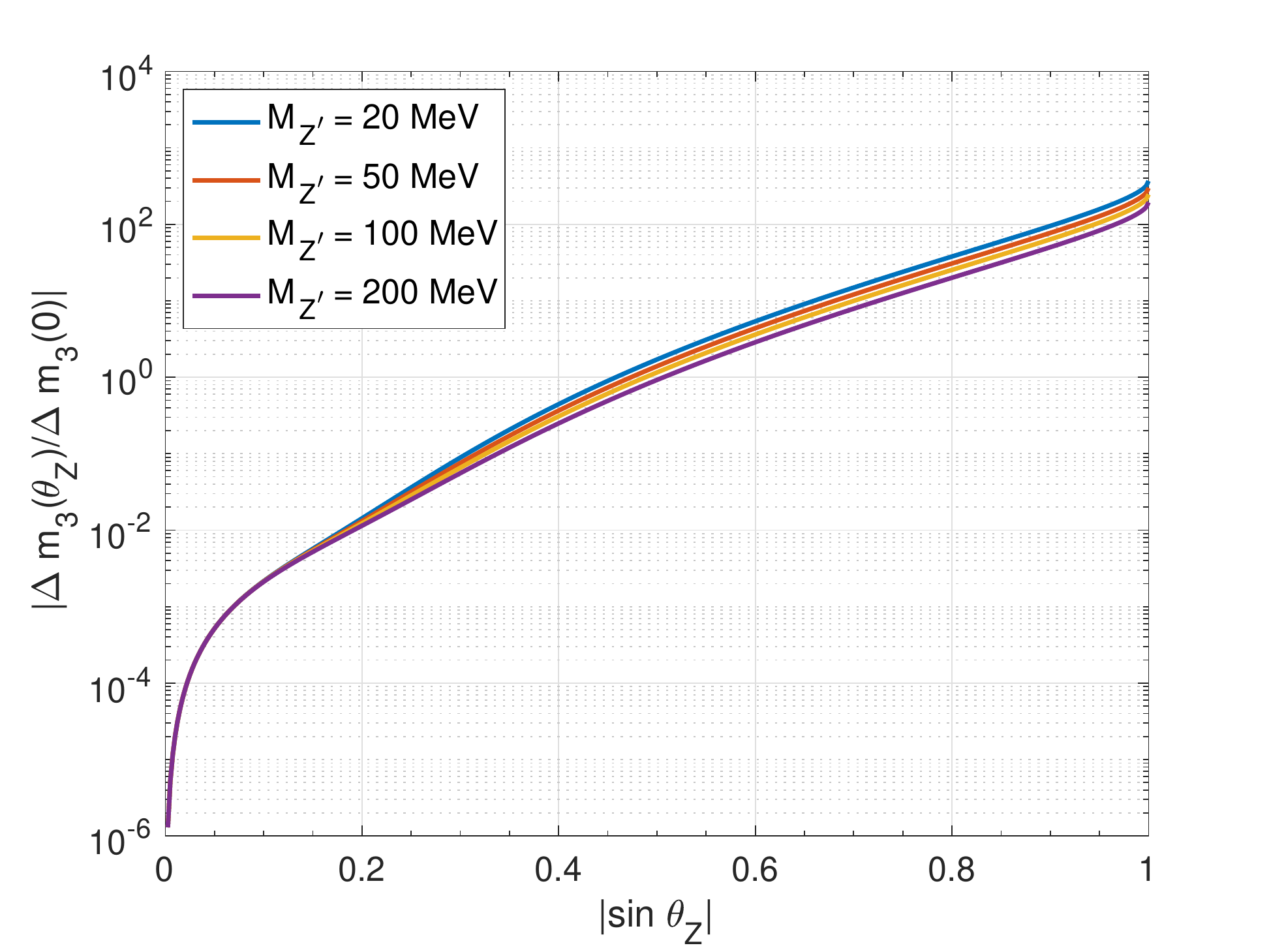}
        \end{center}
        \caption{\label{fig:loopcorrection}
            One-loop corrections to the mass of $\nu_3$, corresponding to
            \textbf{BP1} (see Sect.~\ref{Sec6} for precise definition), with 
            sgn$(\tZ) = +1$, $\Ms = 250$\,GeV and
            $\sin \theta_S = 0.1$. Top: relative correction to $\nu_3$ mass.
            Bottom: correction relative to case $\tZ = 0$.}
    \end{figure}
    
    The correction to the mass of $\nu_3$ (the heaviest of the light neutrinos, 
    assuming  normal mass hierarchy $m_1 < m_2 < m_3$, and obtained by 
    diagonalization of the corrected mass matrix $m_\nu+\delta m_\nu$) due to 
    only the $Z'$ boson is shown in Fig.~\ref{fig:loopcorrection}, denoted by 
    $\Delta m_3$. Looking at the figure, it is clear that the one-loop correction 
    is tiny compared to 
    the mass of $\nu_3$ with our choice of $|\sin \tZ| \ll 1$, relevant to the 
    super-weak case. We consider the mass range $\MZp\in[20,200]$\,MeV where the 
    lightest right-handed neutrino may be sufficiently abundant to provide the 
    correct dark matter energy density with freeze-out scenario 
    \cite{Iwamoto:2021fup}. The smaller $\MZp$ the larger correction.
    
\begin{figure}[ht!]
        \centering
        \includegraphics[width=0.7\linewidth]{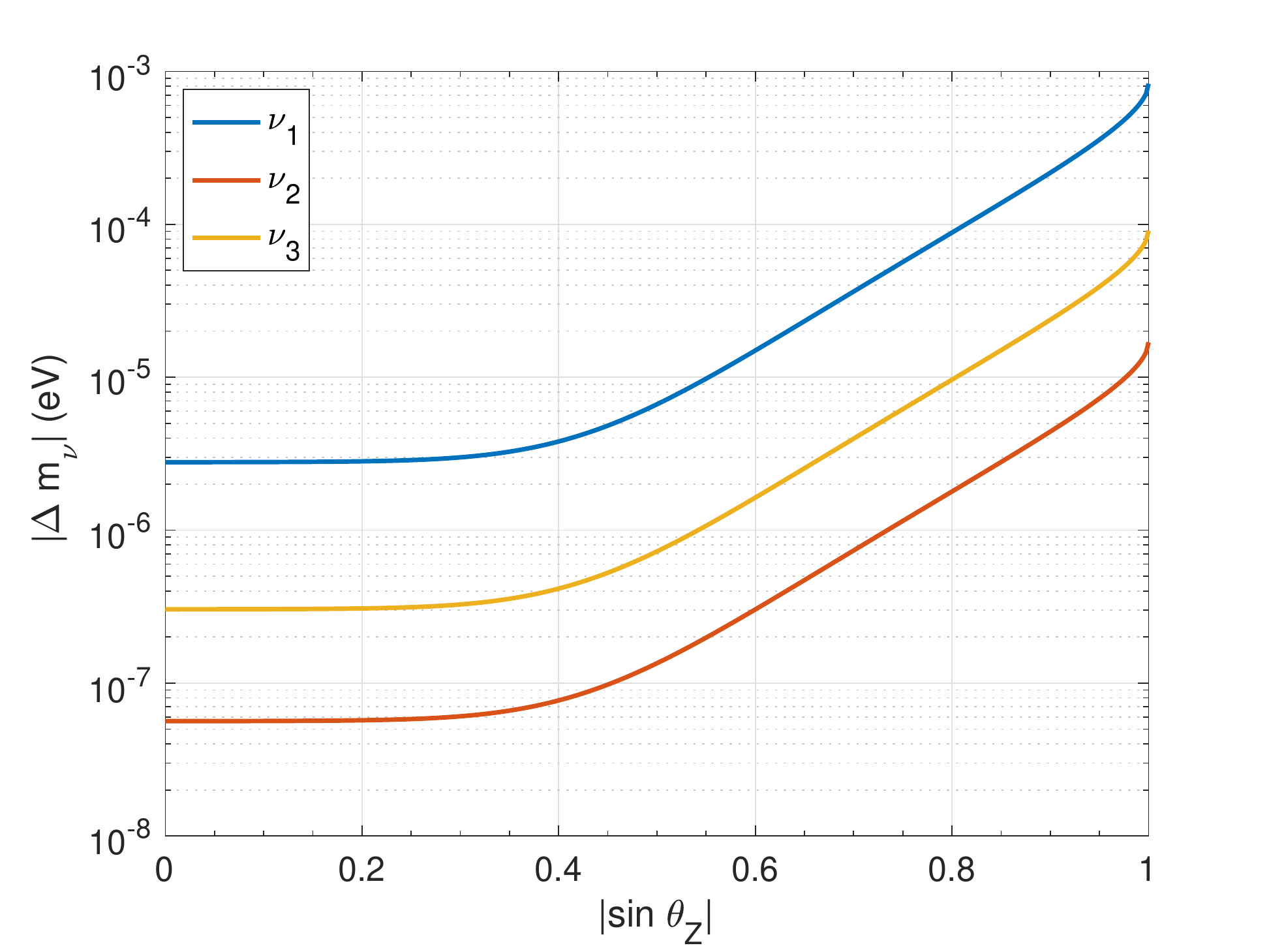}
        \includegraphics[width=0.7\linewidth]{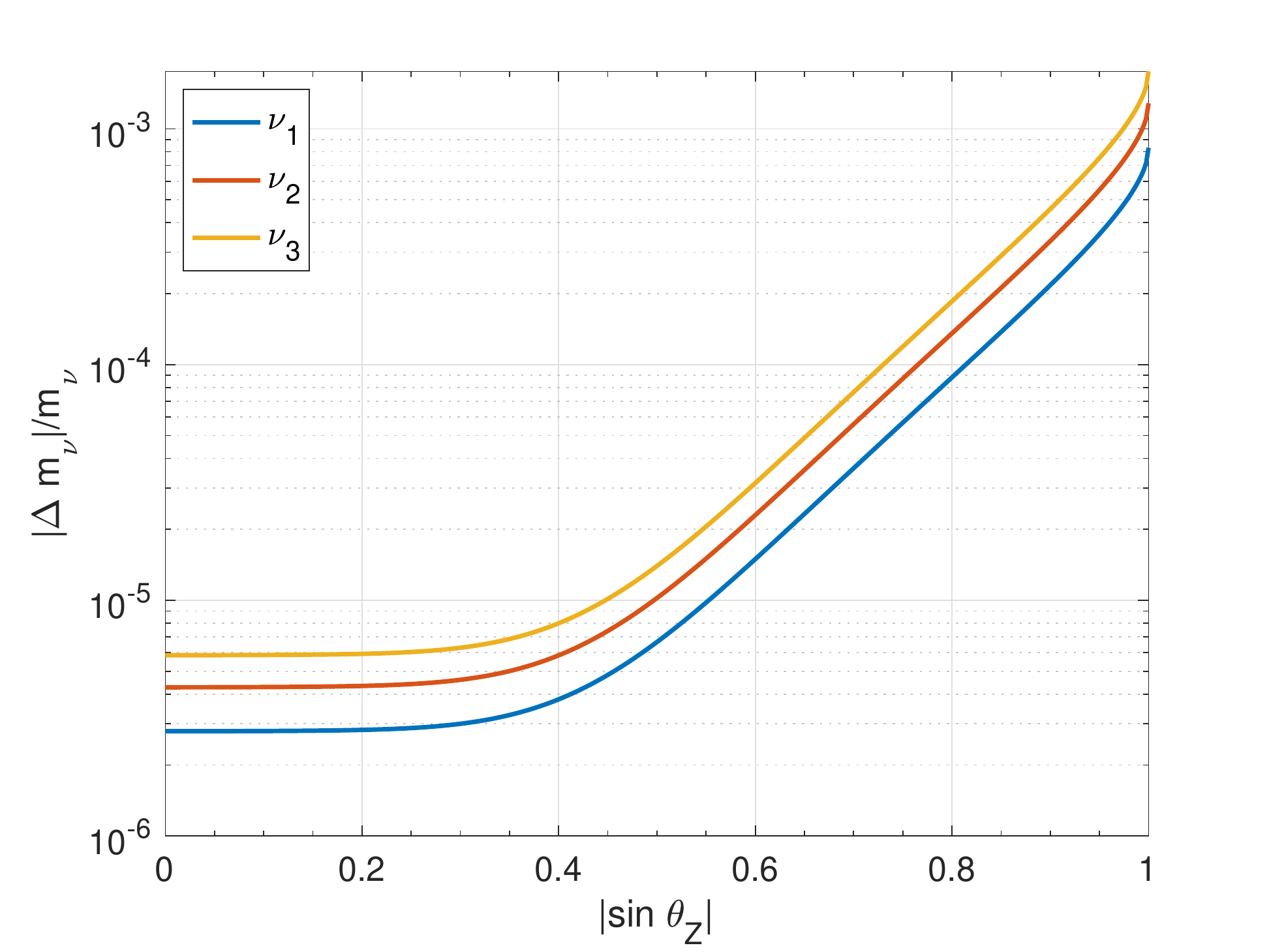}
        \caption{
            \label{fig:loopcorrection2}
            One-loop mass corrections to active neutrinos, for \textbf{BP1}, with
            $\MZp = 50$\,MeV, $\Ms = 250$\,GeV, $\sin \theta_S = 0.1$ and
            sgn$(\tZ) = +1$. Top: Correction in eV units. Bottom: Correction
            relative to tree mass. The plot is almost identical for sgn$(\tZ) = -1$.
        }
    \end{figure}

    In the lower plot of Fig.~\ref{fig:loopcorrection} we show the full 
    correction as a function of $\theta_Z$, $\Delta m_3(\theta_Z)$, 
    relative to the case $\theta_Z = 0$, which corresponds to the $Z'$ boson 
    playing no role in neutrino interactions. It demonstrates how the correction 
    $\Delta m_3$ is dominated by the $Z'$ loop only when $|\sin \tZ|\gtrsim 0.5$. 
    Fig.~\ref{fig:loopcorrection2} shows that the shapes of the plots are
    similar for $\nu_1$ and $\nu_2$, so we can make the same conclusions
    for those neutrinos. 
    
\begin{figure}[t]
        \centering
        \includegraphics[width=0.75\linewidth]{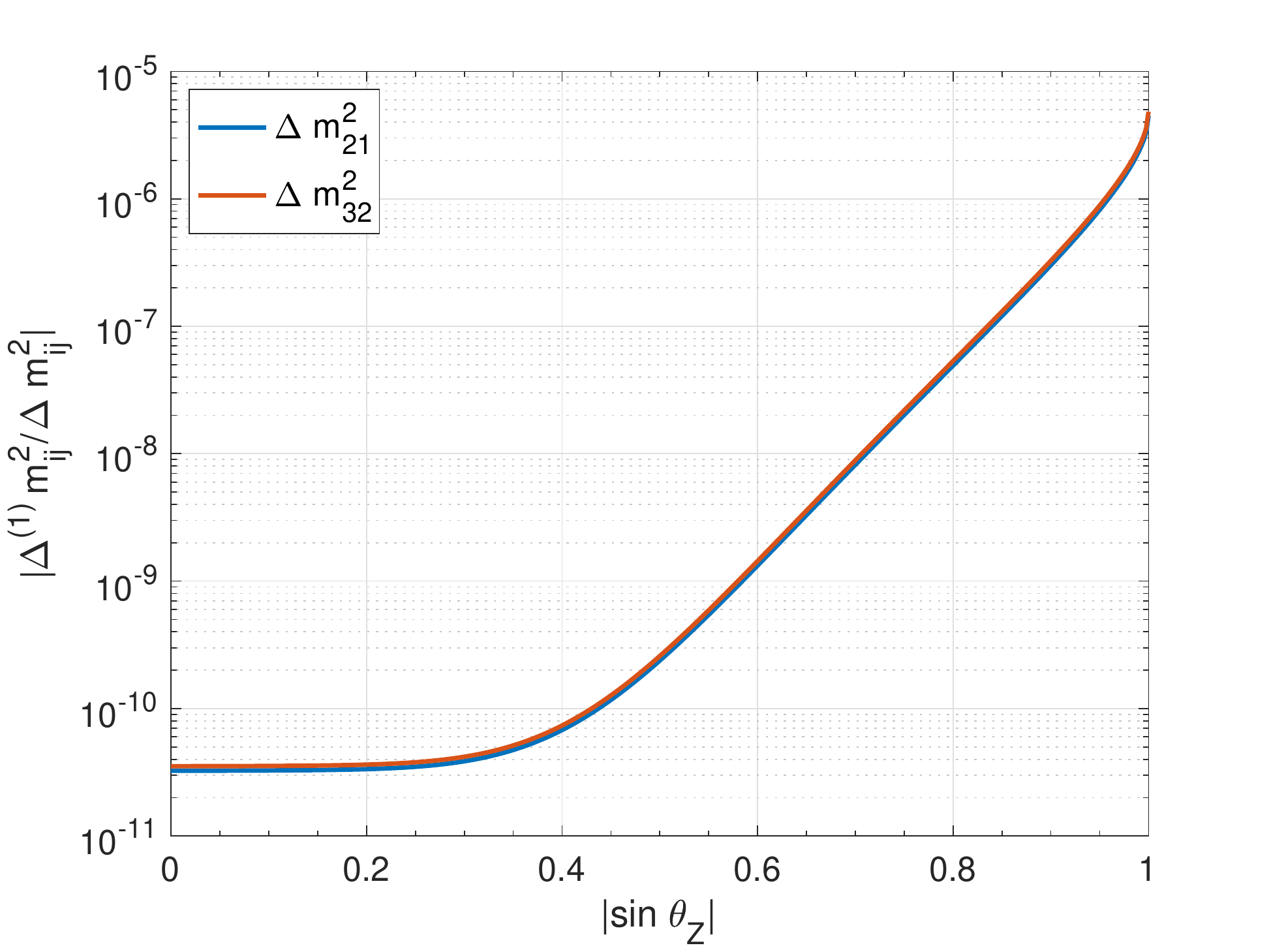}
        \caption{
            \label{fig:osc} Relative one-loop corrections to neutrino oscillation
            parameters $\Delta m_{21}^2$ and $\Delta m_{31}^2$ with respect to
            tree-level parameter, for \textbf{BP1}, with $\MZp = 50$\,MeV,
            $\Ms = 250$\,GeV, $\sin \theta_S = 0.1$ and sgn$(\tZ) = +1$.
        }
    \end{figure}

\begin{figure}[h]
        \centering
        \includegraphics[width=0.7\linewidth]{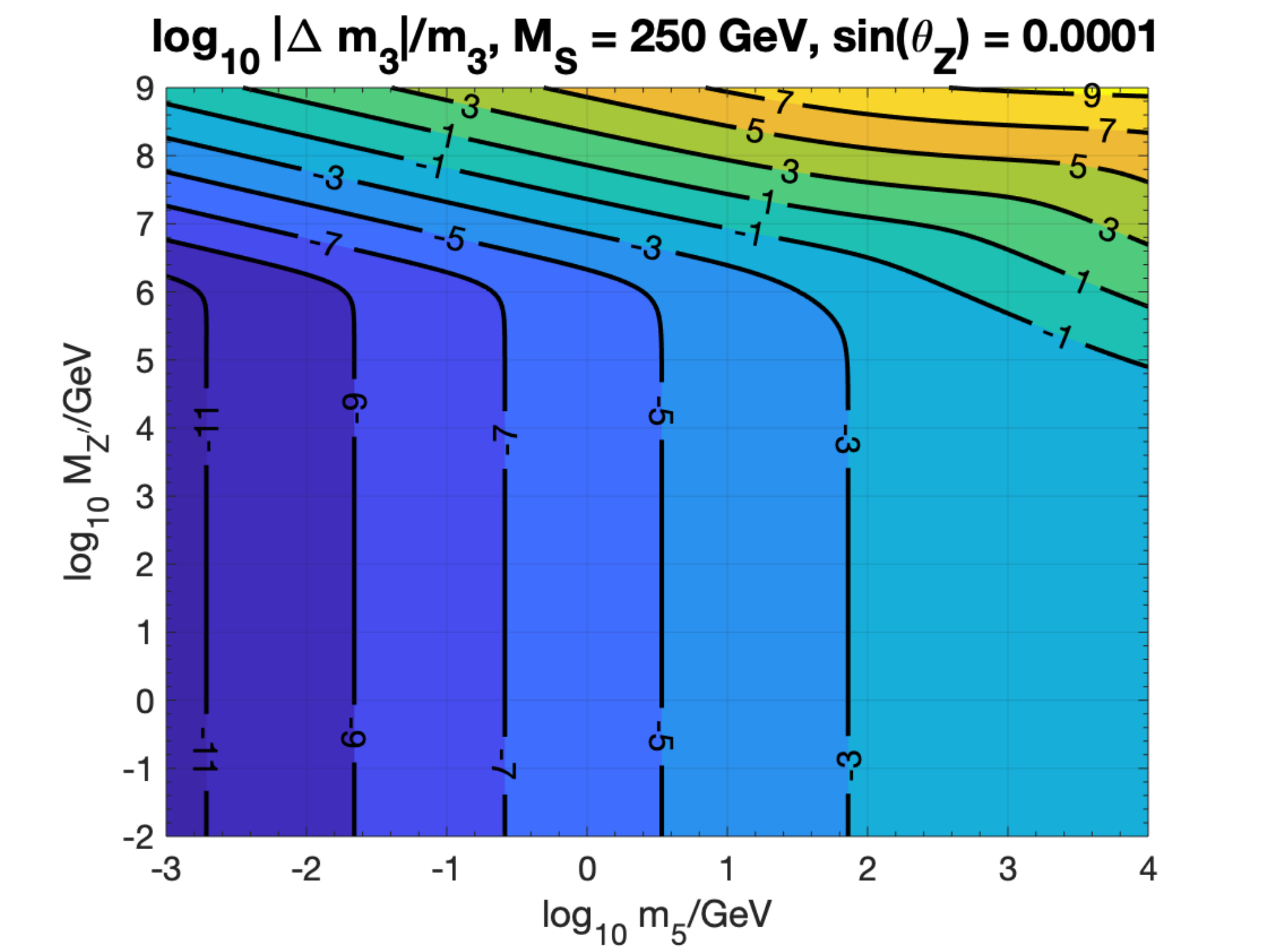}
        \caption{Relative one-loop corrections to $m_3$ in logarithmic
            $(m_5,\MZp)$ plane, for \textbf{BP1} with $\Ms = 250$ GeV, $\sin \theta_S = 0.1$ and $\sin \tZ = 10^{-4}$. Contour labels $n$ correspond to $|\Delta m_3| = 10^n m_3$.
            \label{fig:MNMT}}
    \end{figure}

    The effect of loop corrections on the neutrino oscillation
    parameters is similarly small, and the relative correction for the
    squared mass differences is approximately the mass correction squared,
    as can be seen from Fig.~\ref{fig:osc}. Fig.~\ref{fig:MNMT} shows the 
    contours of relative one-loop corrections to the mass of $\nu_3$ as a
    function of the masses of the $N_2$ neutrino 
    (recall that $m_5 \approx m_6$) and $Z'$. We see that 
    for $m_5 < 100$\,GeV and $\MZp < 10^5$\,GeV the loop corrections are very small.
    
    \subsection{Parametrization of the mass matrix}
    
    Having established the validity of the mass matrix \eqref{eq:fulldiag},
    we use the Casas-Ibarra parameterization \cite{Casas:2001sr} to write the diagonalized light neutrino mass matrix as
    \be \label{light-masses}
    m_\nu^\text{diag} = U_2^T m_\nu U_2 = -U_2^T m_\rD m_\rR^{-1}m_\rD^TU_2 =
    -\frac{v^2}{2} U_2^T Y_\nu m_\rR^{-1} Y_\nu^TU_2.
    \ee 
    Introducing the matrix
    \be \label{r-matrix}
    R = \ri\frac{v}{\sqrt{2}} m_\rR^{-1/2} Y_\nu^T U_2
    (m_\nu^\text{diag})^{-1/2}
    \,,
    \ee 
    we obtain
    \be
    R^T R = -\frac{v^2}{2}(m_\nu^\text{diag})^{-1/2}
    \left( U_2^T Y_\nu m_\rR^{-1/2} m_\rR^{-1/2}Y_\nu^T U_2 \right)
    (m_\nu^\text{diag})^{-1/2}\,,
    \ee 
    which equals the unit matrix because the expression in the parenthesis
    can be simplified using Eq.~(\ref{light-masses}). The relation $R^T R=I$
    is fulfilled by any orthogonal matrix. We can solve Eq.~(\ref{r-matrix}) 
    for the neutrino Yukawa matrix, and find
    \be  \label{casas-ibarra}
    %Y_\nu^T = -\ri \frac{\sqrt{2}}{v}\sqrt{m_\rR}R\sqrt{m_\nu^\text{diag}}U_2^{-1}
    Y_\nu = \frac{\sqrt{2}}{v} U_2^* (m_\nu^\text{diag})^{1/2}(-\ri R^T) m_\rR^{1/2}
    \,.
    \ee 
    
    The inclusion of sterile neutrinos results in non-unitary active-light neutrino 
    mixing matrix \cite{FernandezMartinez:2007ms, Blennow:2011vn} that we write as $(I-\alpha) U_{\rm PMNS}$ where the matrix $\alpha$ is
    proportional to the active-sterile mixing squared, which is tiny. Hence, we
    neglect $\alpha$ in this study, so the active-light mixing is described by the
    unitary matrix $U_{\rm PMNS}= U_{\ell \rL}^\dagger U_2$ (Pontecorvo-Maki-Nakagawa-Sakata 
    matrix). We may choose to set $U_{\ell \rL} = I$, leading to $U_\text{PMNS} = U_2$,
    which is possible if we assume that the charged lepton Yukawa matrix $Y_\ell'$ 
    is invertible%
    \footnote{Non-invertible matrices are rare in the sense that the
        measure of the set of given size such matrices is zero.}, so the
    right-diagonalizing matrix is $U_{\ell \rR} = (Y_\ell^{\prime})^{-1}Y_\ell$. 
    This choice ensures that for the charged leptons the flavour and mass 
    eigenstates coincide. For the active neutrinos the same choice is not
    possible. The $U_2$ PMNS matrix may also include the CP violating and 
    the unknown, complex Majorana phases, but we set those to zero in this study, as we do 
    not expect that such phases will change our conclusions significantly.
    
    % According to Eq.~(\ref{lepton-yukawa}), given the mass spectrum of 
    % light and heavy neutrinos plus the neutrino mixing matrix, the
    % Dirac neutrino mass term related Yukawa matrix $Y_\nu$ can be
    % determined. It can be complex because the PMNS matrix can be complex.
    % The super-weak model does not give predictions on the mixing parameters
    % and on the CP violating phase.  In this study, we assume that the CP 
    % violating phase is zero as the non-vanishing phase would affect our
    % predictions negligibly.
    
    Using that $m_\rR$ is real and diagonal, we can write the mixing 
    matrix \eqref{eq:Uas} as
    \be \label{eq:Uas-2}
    U_{\rm as} = 
    \frac{v}{\sqrt{2}}Y_\nu^* m_\rR^{-1}
    \,.
    \ee 
    We substitute the matrix $Y_\nu$ as given in Eq.~(\ref{casas-ibarra}) 
    to obtain
    \be \label{eq:Uas-3}
    U_{\rm as} 
    = U_\text{PMNS} \sqrt{m_\nu^\text{diag}}(\ri R^\da) m_\rR^{-1/2}
    \,.
    \ee 
    We see that even though the light and heavy neutrino masses and PMNS
    matrix are independent of the choice of $R$ matrix, the mixing between
    active and sterile neutrinos is not so.  One needs to scan over the
    possible orthogonal matrices $R$ to obtain suitable values of
    active-sterile mixing.  In this exploratory study we consider real $R$ matrices, and parameterize it with three Euler angles, whose sines we denote with $s_{12}$, $s_{13}$ and
    $s_{23} \in [0,1]$ and cosines with $c_{ij} \Big(= \sqrt{1-s_{ij}^2}\Big)$:
    \be 
    R(s_{12}, s_{13}, s_{23}) = \left( \begin{array}{ccc}
        c_{12}c_{13} & s_{12}c_{13} & s_{13} \\
        -s_{12}c_{23}-c_{12}s_{23}s_{13} & c_{12}c_{23}-s_{12}s_{23}s_{13} & s_{23}c_{13} \\
        s_{12}s_{23}-c_{12}c_{23}s_{13} & -c_{12}s_{23}-s_{12}c_{23}s_{13} & c_{23}c_{13}
    \end{array}\right) .
    \ee 
    
    \section{Nonstandard interactions\label{Sec5}}
    
\begin{figure}[t!]
        \begin{center}
            \includegraphics[width=0.25\linewidth]{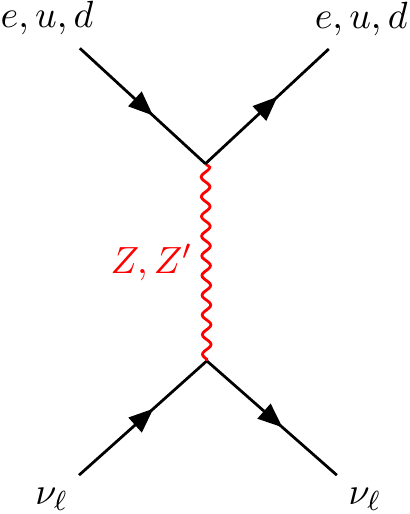}
        \end{center}
        \caption{\label{fig:nsi}
            Neutral gauge boson exchange Feynman diagram. The $Z'$ boson
            contribution is the origin of the NSI in the super-weak model.}
    \end{figure}

    As the neutral $Z'$ boson couples to both active neutrinos and charged
    fermions, it will lead to nonstandard neutrino interactions (NSI), see
    Fig.~\ref{fig:nsi} for a relevant Feynman diagram. Integrating out the
    $Z'$ boson, we find the non-renormalizable dimension-6 effective
    operator \cite{Grossman:1995wx}
    \be \label{NSI}
    \lt_\text{NSI} = -2\sqrt{2}G_{\rm F}\varepsilon_{\ell\ell'}^{ff',X}
    (\overline{\nu_\ell} \gamma^\mu P_\rL\nu_{\ell'})
    (\overline{f'}\gamma_\mu P_X f)
    \ee 
    that in general distorts the neutrino oscillation probabilities. 
    In Eq.~(\ref{NSI}) $\ell$ and $\ell' =$ e, $\mu$, $\tau$
    denote flavour indices, $f$ and $f'$ are any fermions, $P_X$ ($X=$ L, R)
    are chiral projection operators and $G_{\rm F}$ is Fermi's constant. 
    Summation over flavours, fermions and chiralities is implicitly understood. 
    If $f=f'$, we denote $\varepsilon_{\ell\ell'}^{ff,X} = \varepsilon_{\ell\ell'}^{f,X}$.
    The NSI parameters $\varepsilon_{\ell\ell'}^{ff',X}$ and $\varepsilon_{\ell\ell'}^{f,X}$ are dimensionless, and
    in general can be complex numbers, but when they emerge from the
    super-weak force they are real. Given that $Z'$ is light, the mass
    suppression will not ensure that the NSI is small.  Instead, the
    suppression is due to the small couplings.
    
    Man-made neutrinos are utilized in oscillation experiments.  Neutrinos
    propagate through Earth matter, consisting of electrons, protons and
    neutrons (i.e.~u and d quarks).  $Z'$-mediated $\nu \ell \to \nu\ell$
    and $\nu q \to \nu q$ scatterings will produce an extra term in the
    neutral current Mikheyev-Smirnov-Wolfenstein potential
    \cite{Wolfenstein:1977ue,Mikheev:1986gs,Mikheev:1986wj}, 
    which is responsible for the matter effects in neutrino oscillations.
    However, the $Z'$ boson couples to neutrino flavours universally. 
    If there were no sterile neutrinos, this would not affect oscillation 
    probabilities, but in super-weak model this is not the case. As the full 
    light neutrino mixing matrix $(I-\alpha)U_\text{PMNS}$ is not unitary, the
    NSI contribution to the neutrino oscillation transition probabilities 
    will be nonzero, but suppressed by the small absolute values of the 
    elements of active-sterile mixing matrix $U_{\rm as}$. An oscillation 
    experiment probing matter NSI parameters will measure only effective NSI,
    $\varepsilon^{\rm m, eff} \sim \varepsilon^{\rm m}U_{\rm as}$,
    when both NSI and nonunitary contributions are present \cite{Blennow:2016jkn}. 
    We see from Eq.~\eqref{eq:Uas-2} that for the keV, MeV and GeV scale neutrinos 
    that we are
    considering, the absolute values of the elements of matrix $U_{\rm as}$ are much
    smaller than the present and near-future experimental limits on the NSI parameters. 
    Thus we conclude that the near-future neutrino oscillation experiments will not
    be sensitive enough to probe NSI originating from the super-weak model 
    (see the Appendix for more details). 
    
    The $Z'$ boson retains the lepton universality also for charged leptons,
    so it will not contribute to any charged lepton flavour violating
    processes. The super-weak force does not affect the charged
    currents of the SM, so there will not be any charged current NSI
    operators. Also, the super-weak force does not imply the existence of
    flavour-changing neutral currents. Nevertheless, the existence of the $Z'$ boson
    will contribute to the effective mass of the neutrinos in matter and to neutrino-electron
    and neutrino-quark scattering processes. In fact, in neutrino scattering experiments the
    NSI are not suppressed by active-sterile mixing, thus those are more sensitive to the
    NSI as compared to oscillation experiments, and provide the primary constraints for the
    super-weak model in the neutrino sector. In what follows, we explore those constraints.
    
    \subsection{Nonstandard interactions in medium}
    
    The matter NSI is obtained by integrating out the $Z'$ boson mediator
    in the elastic $\nu_\ell f \to \nu_\ell f$ scattering amplitude,
    leading to the effective Lagrangian,
    \be 
    \lt_{\rm NSI} = -\frac{1}{\MZp^2}(eC_{Z'\nu_\ell\nu_\ell}^\rL)
    (eC^X_{Z'ff})(\overline{\nu_\ell} \gamma^\mu P_\rL \nu_\ell)
    (\overline{f}\gamma_\mu P_Xf) 
    \,.
    \label{eq:Deltalt}
    \ee
    Matching Eq.~(\ref{eq:Deltalt}) to Eq.~(\ref{NSI}) reveals the form of
    the effective couplings:
    \be 
    \varepsilon^{f,X}_{\ell\ell} = \frac{v^2}{2\MZp^2}(eC_{Z'\nu_\ell\nu_\ell}^\rL)(eC_{Z'ff}^X)
    \,,\quad 
    \varepsilon^{f}\equiv
    \varepsilon^{f}_{\ell\ell}=
    \varepsilon^{f,\rL}_{\ell\ell} + \varepsilon^{f,\rR}_{\ell\ell}
    \ee 
    where we indicated that in the super-weak model the NSI coupling is the
    same for all neutrino flavours:
    $\varepsilon^f_{ee} = \varepsilon^f_{\mu\mu} = \varepsilon^f_{\tau\tau}$. 
    In principle, flavour non-universal couplings are also possible, and would cause
    significantly stronger oscillation-distorting NSI and charged lepton violating
    decays,  but we do not discuss such an option here.
    
    As matter is electrically neutral and consists only of electrons,
    protons and neutrons, summing all contributions gives
    \be
    \bsp
    \varepsilon^{\rm m} &= \varepsilon^{e} +2\varepsilon^{u}
    +\varepsilon^{d} +\frac{N_{\rm n}}{N_\re}
    (\varepsilon^{u}+2\varepsilon^{d})\\
    &= -\frac{v^2 }{8 \MZp^2}\frac{N_{\rm n}}{N_\re}
    \left(g_y' \cos \tZ - \frac{g_\rL \sin \tZ }{\cos \theta_W}\right)
    \left((g_y'-g_z') \cos \tZ-\frac{g_\rL \sin \tZ }{\cos \theta_W}\right)
    \esp
    \ee
    where $N_{\rm n}$ and $N_\re$ are the neutron and electron number densities.  
    Interestingly, the electron and proton contributions cancel each other, 
    and the effective coupling is proportional to neutron density in matter.
    
    In the small $\tZ$ limit, we obtain the following approximation:
    \begin{align}
        \varepsilon^{\rm m}
        &\simeq \frac{v^2}{8\MZp^2}\frac{N_{\rm n}}{N_\re}\left[ g_\rL \sin\tZ  \frac{2g_y'-g_z'}{\cos \theta_W}- \frac{g_\rL^2 \sin^2\tZ}{\cos^2 \theta_W}-g_y^{\prime 2}   +g_y' g_z' \right]
    \end{align}
    We can obtain a numerical estimate for the NSI strength by substituting
    $v=246.22$\,GeV, $g_\rL = 0.652$ and $\cos \theta_W = 0.8819$:
    \be
    \bsp
    \varepsilon^{\rm m} &\simeq 0.7578\cdot \frac{N_{\rm n}}{N_\re} 
    \left( \frac{10 \text{ MeV}}{\MZp}\right)^2 
    \\& \times 
    \left(
    \frac{\sin \tZ}{1.353\cdot 10^{-4}}
    \frac{2g_y'-g_z'}{10^{-4}}
    - \left( \frac{ \sin \tZ}{1.353\cdot 10^{-4}}\right)^2
    - \left(\frac{g_y^{\prime}}{10^{-4}}\right)^2
    + \frac{g_y' g_z'}{10^{-8}}
    \right) 
    \esp
    \ee
    
    \begin{figure}[t]
        \begin{center}
            \includegraphics[width=0.49\linewidth]{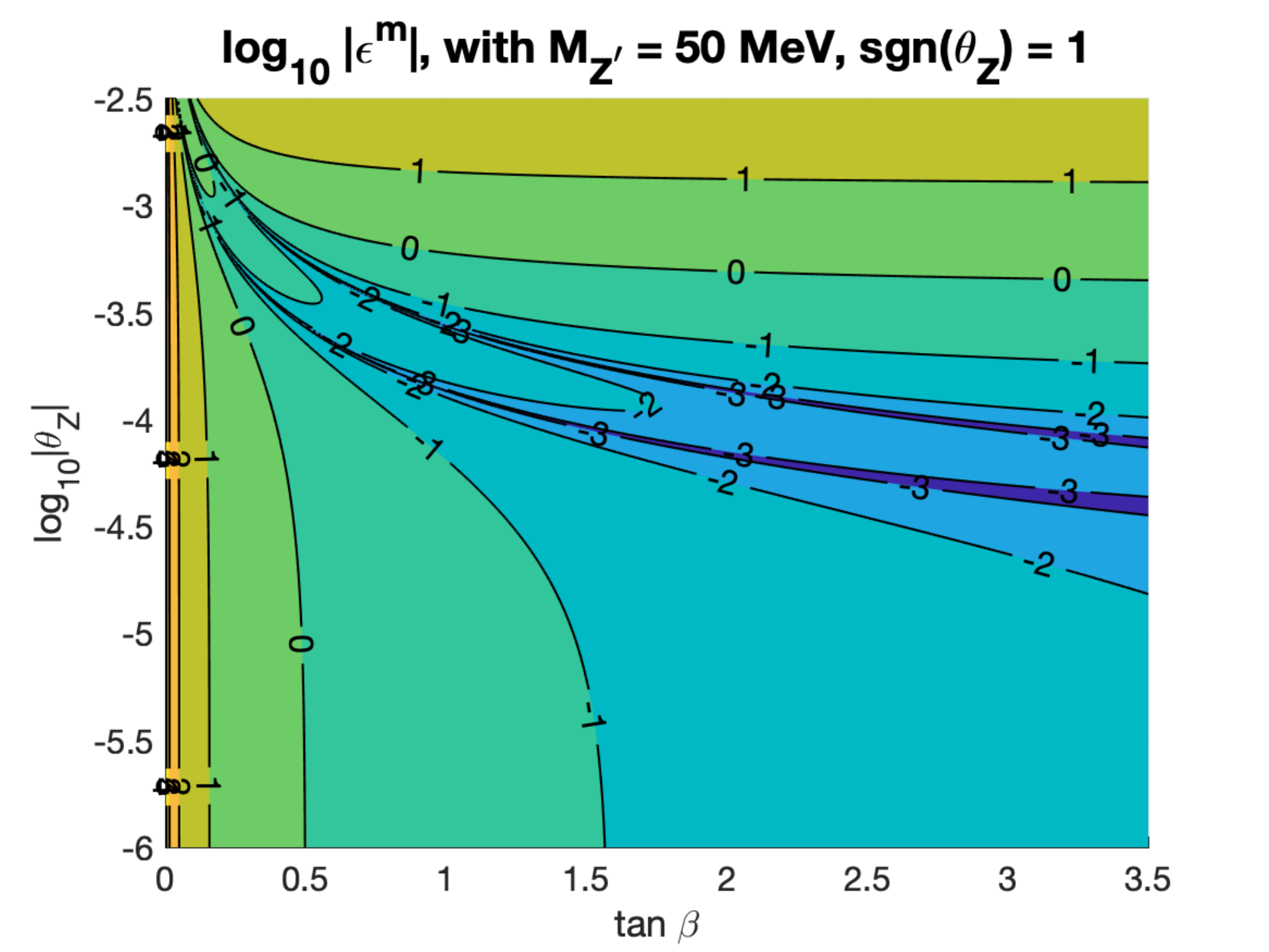}
            \includegraphics[width=0.49\linewidth]{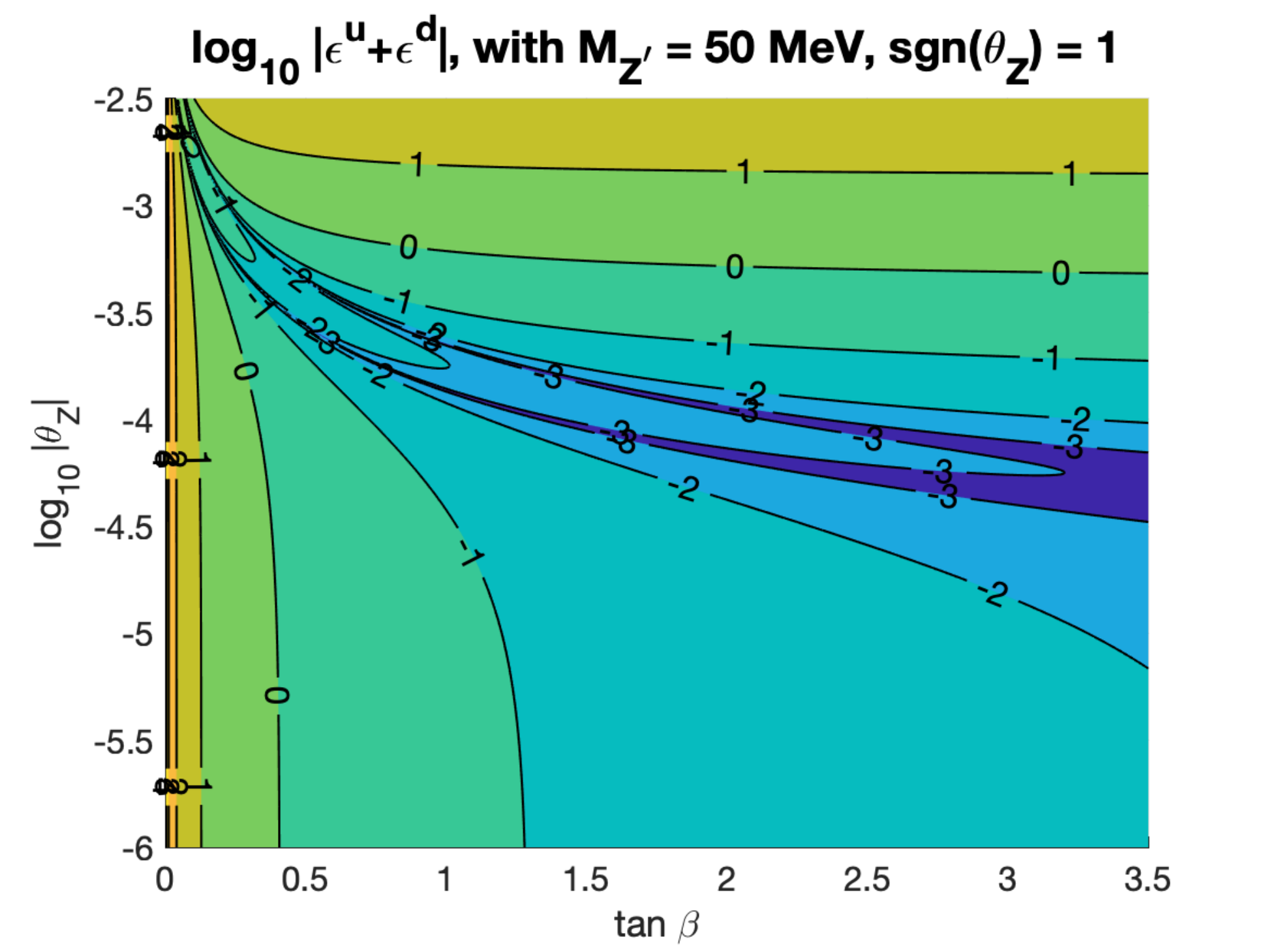}
            \includegraphics[width=0.49\linewidth]{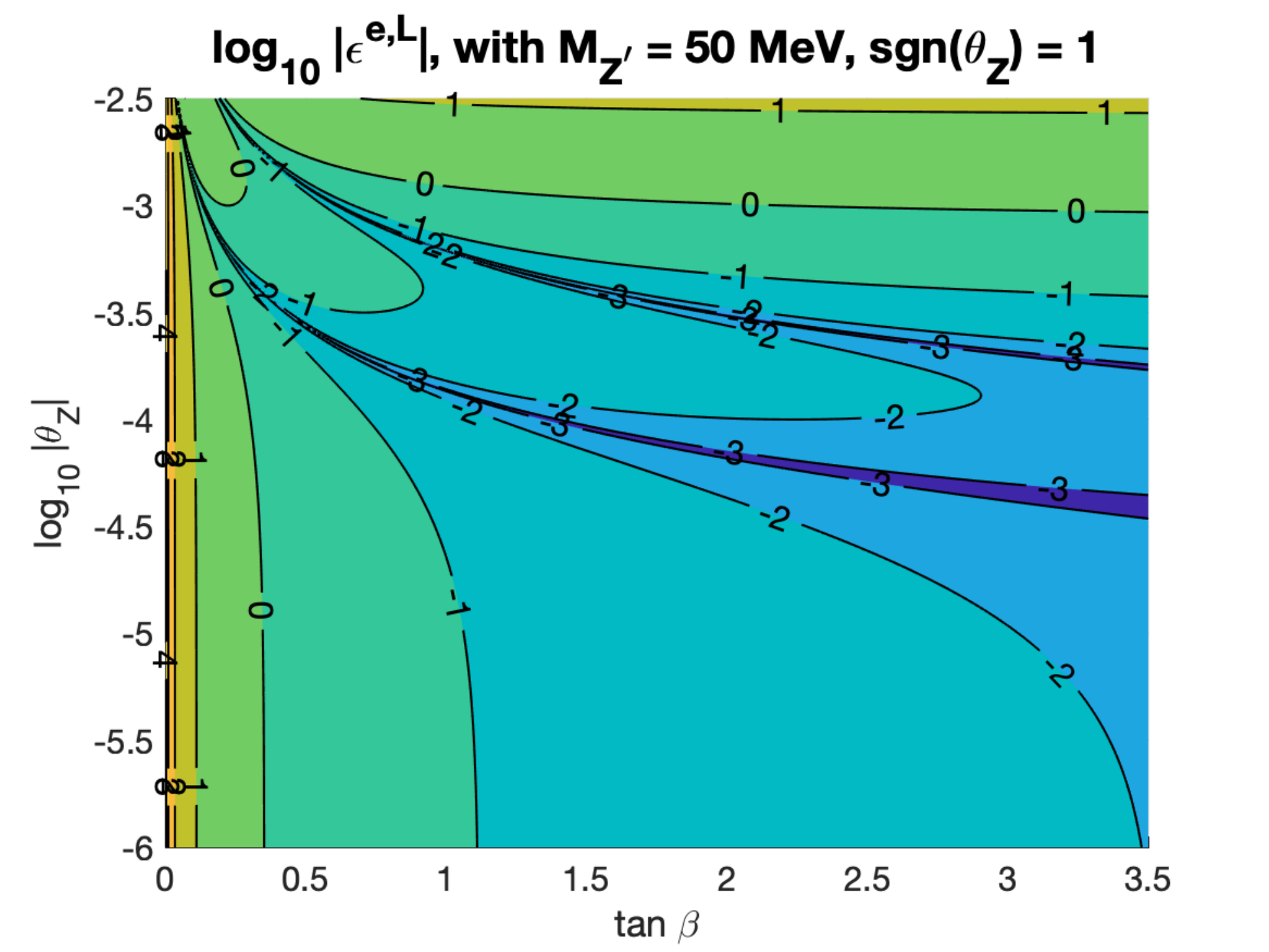}
            \includegraphics[width=0.49\linewidth]{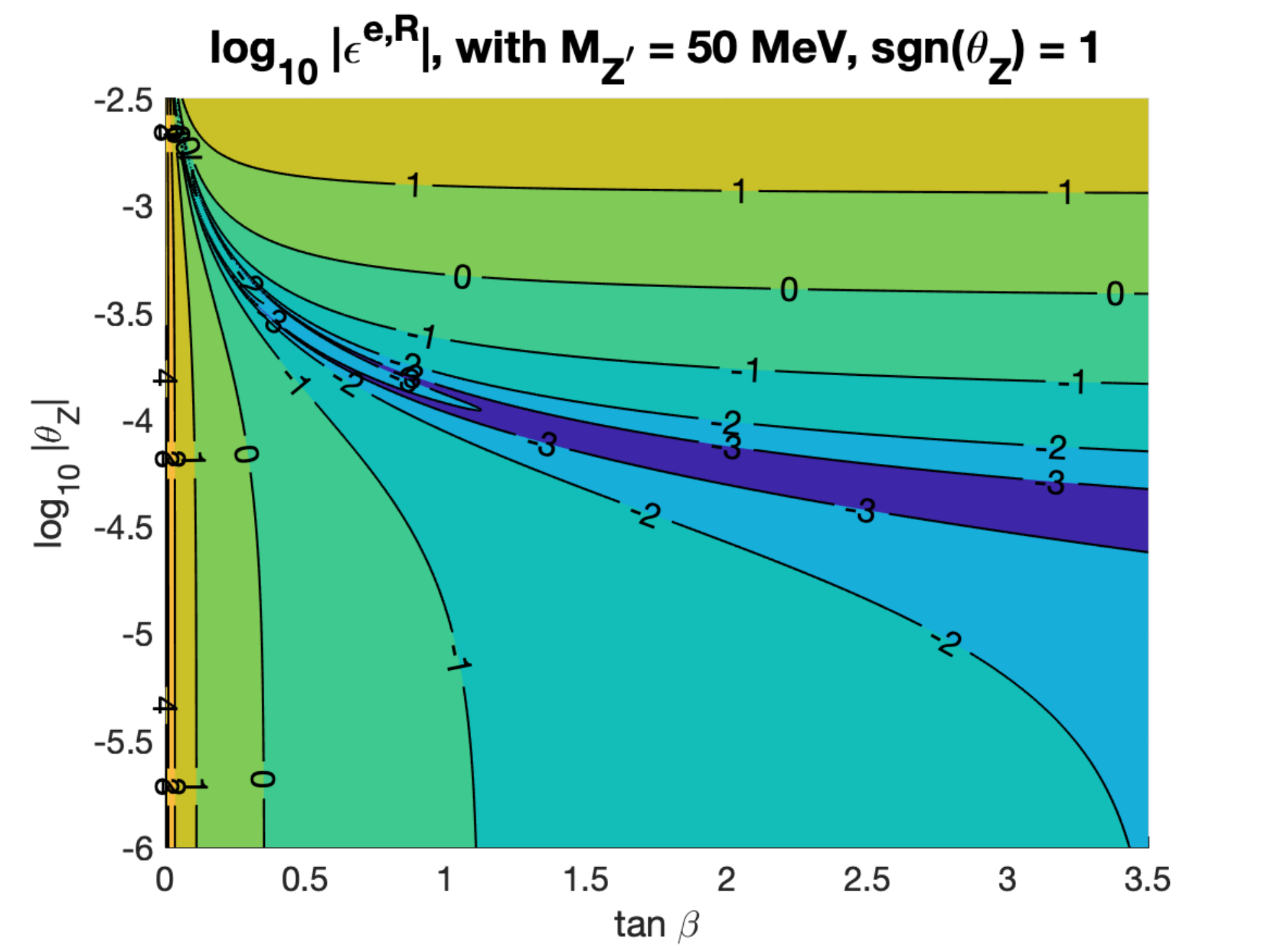}
            \caption{\label{fig:NSI}Contour plots of $\log_{10} |\varepsilon^{f}|$
                for sgn($\tZ) = +1$ in the semilogarithmic ($\tan \beta, \log_{10} |\tZ|$)
                plane with $\MZp = 50$\,MeV. Contour label $n$ corresponds to value
                $\varepsilon^{f} = 10^n$.
                Upper left: total NSI, $\varepsilon^{\rm m}$.
                Upper right: quark NSI, $\varepsilon^{u} +\varepsilon^{d}$.
                Lower left: left-chiral lepton NSI, $\varepsilon^{{e},\rL}_{\ell\ell}$.
                Lower right: right-chiral lepton NSI, $\varepsilon^{{e},\rR}_{\ell\ell}$.}
        \end{center}
    \end{figure}
    %\clearpage
    
    \begin{figure}[t]
        \begin{center}
            \includegraphics[width=0.49\linewidth]{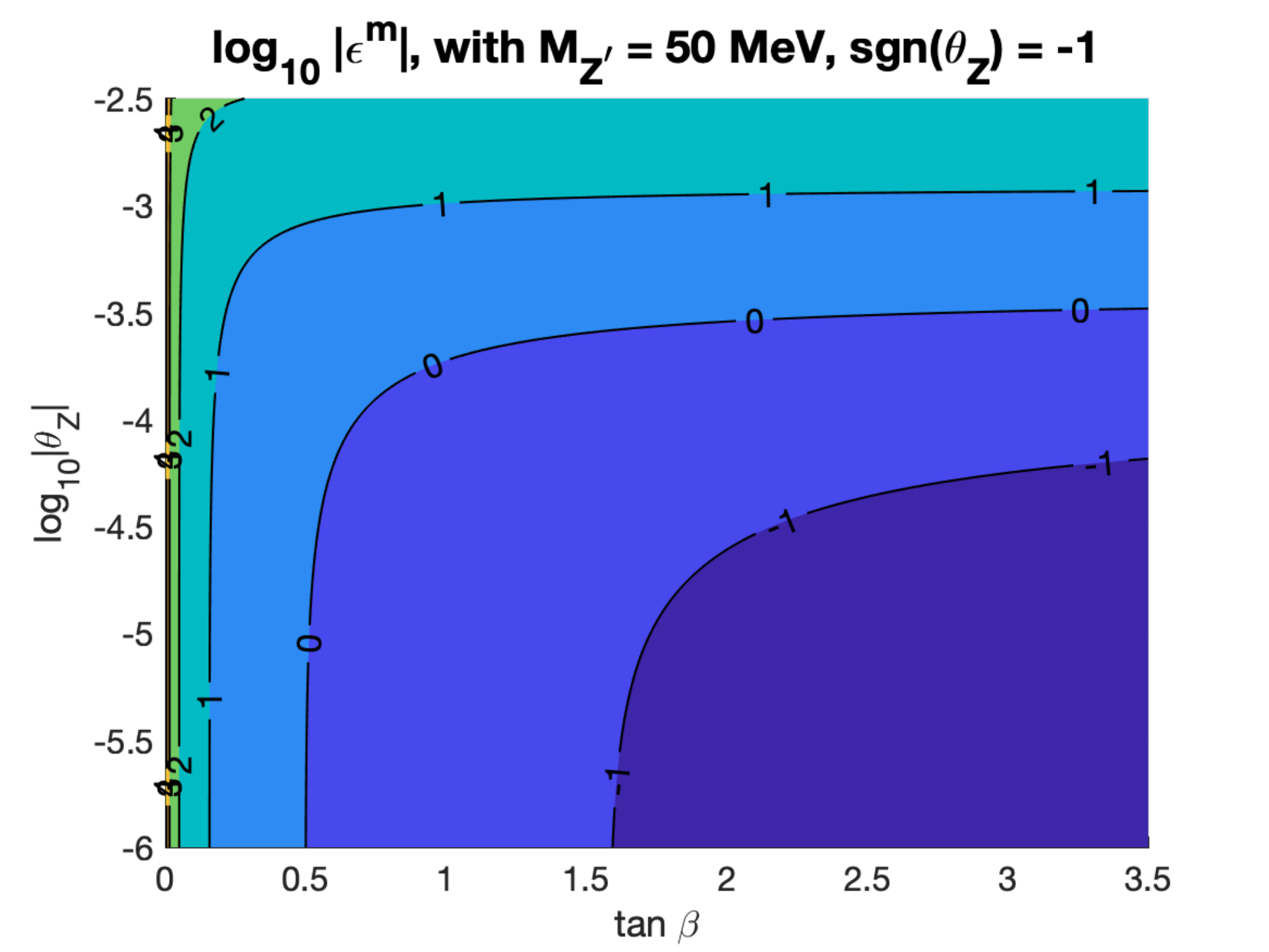}
            \includegraphics[width=0.49\linewidth]{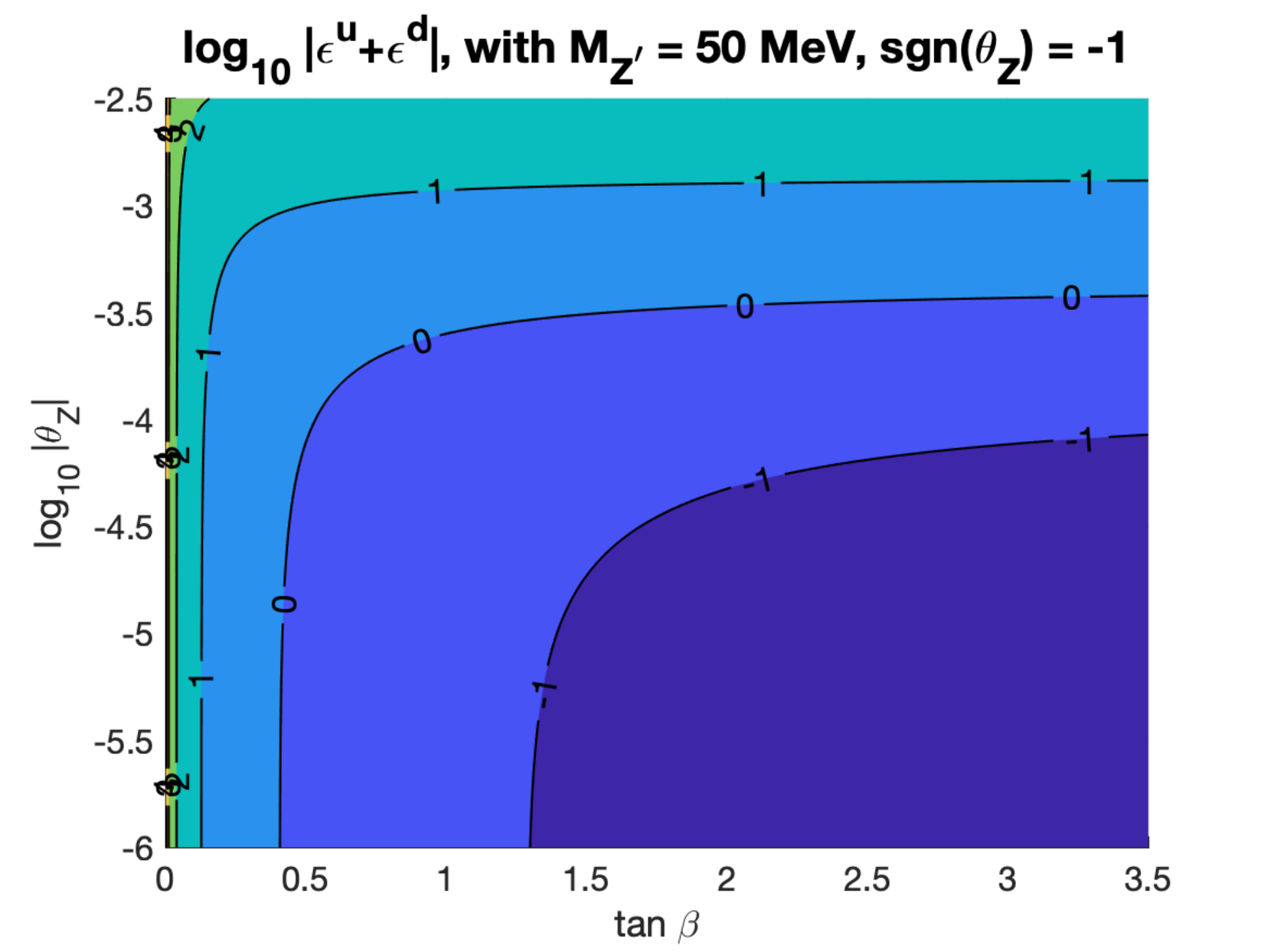}
            \includegraphics[width=0.49\linewidth]{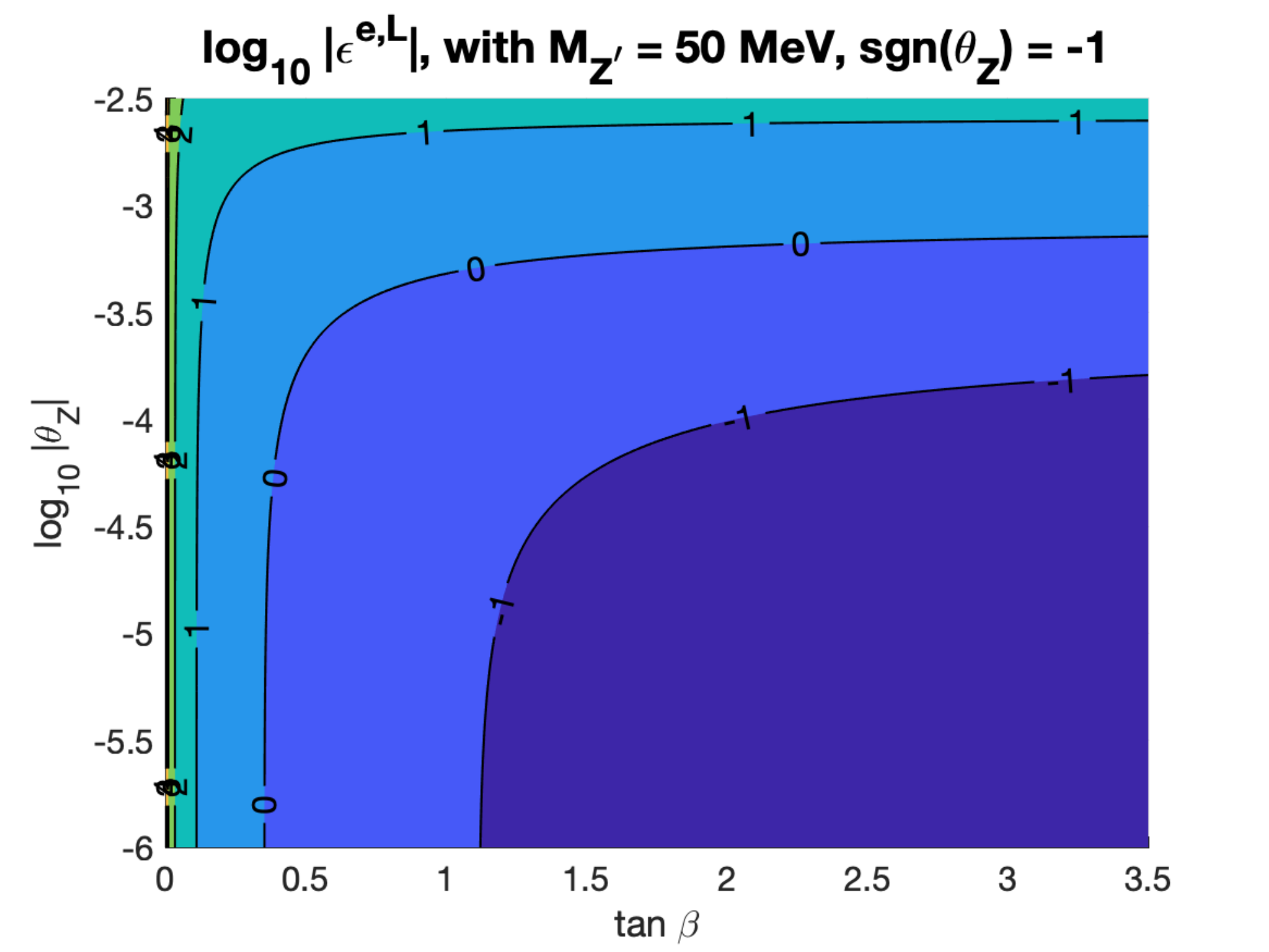}
            \includegraphics[width=0.49\linewidth]{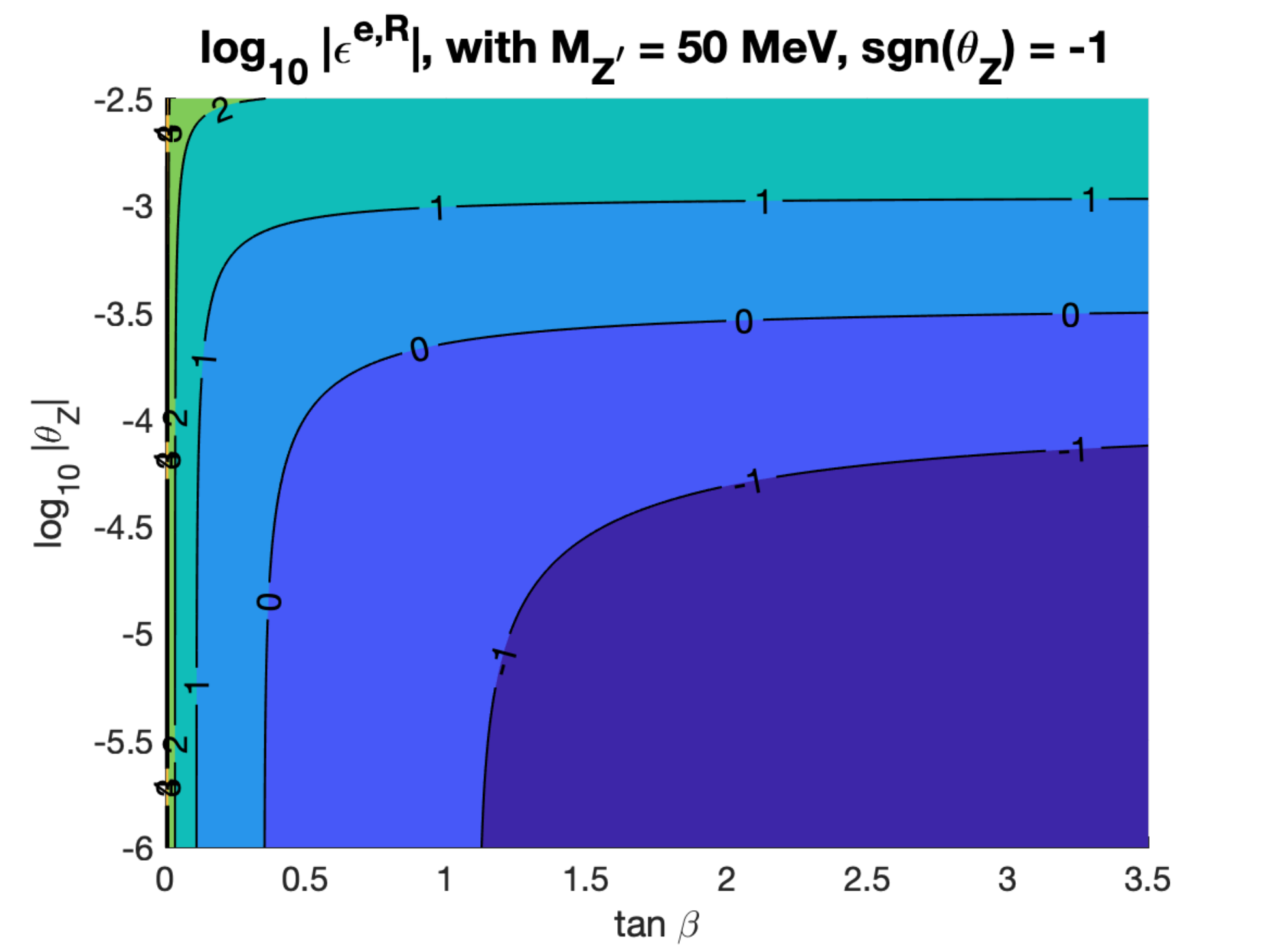}
            \caption{\label{fig:NSI2}Same as Fig.~\ref{fig:NSI}, but for sgn($\tZ) = -1$.}
        \end{center}
    \end{figure}
    
    In Figs.~\ref{fig:NSI} and \ref{fig:NSI2} we demonstrate for a selected
    value of $\MZp$ that it is possible to produce small NSI. We chose
    $\MZp=50$\,MeV, within the range relevant for freeze-out dark matter production
    \cite{Iwamoto:2021fup}. We see that the allowed region (within the $10^{-1}$ 
    contours) is constrained to $-10^{-5} \lesssim \tZ < 10^{-4}$. For $\tZ < 0$,
    small NSI is produced only if $\tan \beta \gtrsim 1.5$. For $\tZ > 0$, 
    there is no constraint for $\tan \beta$, but for $\tan \beta \lesssim 1.5$ 
    there will be also a lower bound on $\tZ$, and the available range for
    $\tZ$ shrinks to a narrow interval as $\tan \beta$ decreases (see Fig.~\ref{fig:NSI}).
    
    \begin{figure}[t]
        \begin{center}
            \includegraphics[width=0.8\linewidth]{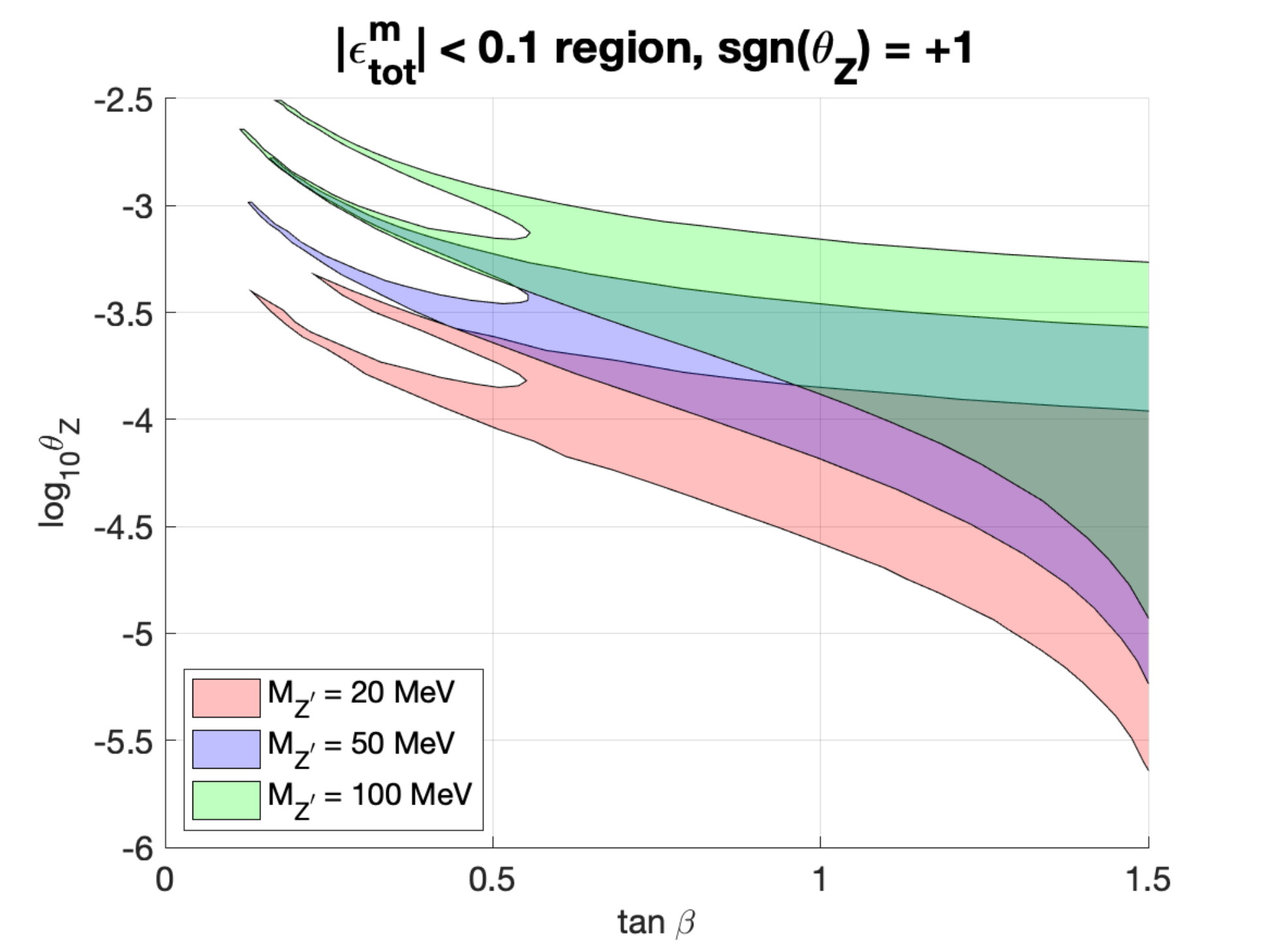}
            \caption{\label{fig:nsicombined}
                The available parameter space consistent with $|\varepsilon^{\rm m}| <
                0.1$ for sgn$(\tZ) = +1$ in semi-logarithmic $(\tan \beta,
                \log_{10}\tZ)$ plane.  }
        \end{center}
    \end{figure} 
    
    Increasing the mass of the $Z'$ boson will decrease the NSI strength,
    and therefore the contours in Fig.~\ref{fig:NSI} and
    Fig.~\ref{fig:NSI2} will move upwards.  We demonstrate this behaviour
    in Fig.~\ref{fig:nsicombined} for the region $|\varepsilon^{\rm m}|
    \lesssim 0.1$. Our choice for this region is motivated by the current
    experimental bounds $|\varepsilon^{e,u,d}_{\ell\ell}| \lesssim 0.1$ \cite{Biggio:2009nt,Ohlsson:2012kf}.
    The NSI constraint forces the gauge couplings $g_y'$ and $g_z'$ be
    quite small, less than about $10^{-4}$.

    We conclude this subsection with a remark. Given a medium with electron
    density $N_\re = 10^{30}$ m$^{-3}$, the effective neutrino mass (for all
    flavours in flavour-universal case) in medium is changed by
    \be 
    \Delta m_\nu^{\rm m} = \sqrt{2}G_{\rm F}N_e\varepsilon^{\rm m} = V_\text{CC}\varepsilon^{\rm m}  \ll m_\nu \sim 0.01\,\text{eV},
    \ee 
    where $V_\text{CC} =\rO(10^{-13})$\,eV is the matter potential
    in the Earth and $\rO(10^{-11})$ eV in the Sun.
    
    \subsection{Changes to the neutrino-electron elastic scattering}
    
    In the super-weak model, the flavour-conserving NSI gives an additional
    contribution to neutrino-electron elastic scattering cross section
    \cite{Davidson:2003ha}: 
    \be 
    \sigma(\nu_e e \rightarrow \nu_e e) =
    \frac{2}{\pi}G_{\rm F}^2m_e E_\nu\left( \Big(1+g_\rL^e + \varepsilon^{e,\rL}_{ee}\Big)^2 + \frac{1}{3}\Big(g_\rR^e+\varepsilon^{e,\rR}_{ee}\Big)^2\right),
    \ee 
    where the NSI couplings are
    \be
    \bsp
    \varepsilon^{e,\rL}_{ee} &= 
    \frac{v^2}{8\MZp^2}\left((g_y'-g_z') \cos \tZ-g_\rL \frac{\sin \tZ}{\cos \theta_W}\right) 
    \\&\qquad\quad\times
    \left((g_y'-g_z') \cos \tZ+g_\rL  \frac{1-2\sin^2 \theta_W}{\cos \theta_W}\sin \tZ\right) \\
    \varepsilon^{e,\rR}_{ee} &= 
    \frac{v^2}{8\MZp^2}\left((g_y'-g_z') \cos \tZ-g_\rL \frac{\sin \tZ}{\cos \theta_W}\right)
    \\&\qquad\quad\times
    \left((2 g_y'-3 g_z') \cos \tZ-2 g_\rL \frac{\sin^2 \theta_W}{\cos \theta_W}\sin \tZ\right) 
    \esp
    \ee
    Comparing these cross sections to results of scattering experiments the upper
    bounds $|\varepsilon^{e,\rL}_{ee}|, |\varepsilon^{e,\rR}_{ee}| \lesssim 0.1$
    are derived in Ref.~\cite{Davidson:2003ha}. 
    
    \section{Benchmark points\label{Sec6}}
    
    We searched for benchmark points in the ranges $m_1 \in [0, 50]$\,meV, 
    $m_4 \in [1,50]$\,keV, except for \textbf{BP5} for which $m_4$ is chosen 
    in the MeV range where $N_1$ can be a proper freeze-out dark matter candidate. 
    The other benchmarks are suitable for the freeze-in case \cite{Iwamoto:2021fup}.
    We have chosen \textbf{BP2} so that it is consistent with the
    unidentified 3.5\,keV X-ray line. 
    We assume almost mass-degenerate sterile neutrinos $N_2$ and $N_3$ with masses
    $m_{5,6} \in [1,80]$\,GeV, motivated by the $\nu$MSM scenario
    \cite{Asaka:2005pn,Shaposhnikov:2006xi}, although it is also possible to 
    relax on this assumption of mass degeneracy. As mentioned in
    Sect.~\ref{Sec3}, there are stringent bounds on active-sterile mixing in this 
    mass range. For the sterile neutrino $N_1$ with $m_4$ in the keV range 
    the corresponding bound is more relaxed, $|U_{a 4}|^2 \lesssim 10^{-4}$
    \cite{Atre:2009rg}. Studies on future tritium beta decay experiments imply 
    that the sterile-$\nu_e$ mixing bound may improve significantly 
    to $|U_{e4}|^2 \lesssim 10^{-8}$ \cite{Mertens:2014nha}. 
    Bounds for keV neutrino mixing can be set from kink searches in the $\beta$
    spectrum of radioactive nuclei, while for MeV neutrinos from peak searches 
    from $\pi$ and $K$ meson decays 
    \cite{Britton:1992pg,Britton:1992xv,Yamazaki:1984sj,Artamonov:2014urb}, 
    and  for GeV neutrinos from collider experiments, 
    with a CMS result \cite{Sirunyan:2018mtv} obtained at the 13 TeV being
    the most constricting for $m_{5,6} > 10$\,GeV. The Future Circular Collider 
    in the lepton collision mode will have the possibility to test a large portion 
    of the parameter space favorable by our benchmark points. 
    
    We used $v = 246.22$\,GeV for the VEV of the scalar field $\phi$ and  $w\in [100,750]$\,GeV for the VEV of the second scalar.  In
    our analysis the mass of $Z'$ boson plays a role only when we calculate
    the one-loop corrections to the light neutrino masses and the
    tree-level contributions to nonstandard interactions.  The values for
    the Yukawa and the mixing matrices are independent of the gauge sector.
    The input values for each benchmark point are given in Table~\ref{BPtable}.
    The one-loop corrections to active
    neutrino masses are dependent also on the gauge parameters $g_y'$,
    $g_z'$ and $\tZ$.  These parameters do not affect any other observables
    in the neutrino sector. A related analysis has been carried out in 
    Ref.~\cite{Ballett:2019cqp} where the new U(1) coupling is of order
    unity, the mass of $Z'$ is at the GeV scale and tree-level neutrino 
    masses vanish.
    \begin{table}[t!]
        \begin{center}
            \caption{\label{BPtable}Input values of benchmark points:
                $R$ matrix paramerers, neutrino masses and
                vacuum expectation value of the new scalar field. We assume normal mass hierarchy $m_1 < m_2 < m_3$}
            \begin{tabular}{|c|c|c|c|c|c|}\hline 
                \textbf{Benchmark point }& \textbf{BP1} & \textbf{BP2} & \textbf{BP3} & \textbf{BP4} & \textbf{BP5}\\ \hline 
                \rule{0pt}{3ex} $s_{12}$       & 0.61   & 0.15 & 0.28 & 0.60 & 0.68 \\\hline 
                \rule{0pt}{3ex} $s_{13}$       & 0.3126 & 0.10 & 0.86 & 0.58 & 0.40\\\hline
                \rule{0pt}{3ex} $s_{23}$       & 0      & 0.40 & 0    & 0.16 & 1.00 \\\hline 
                \rule{0pt}{3ex} $m_1$ (meV)    & 10     & 1    & 0    & 0.1  & 5 \\\hline 
                \rule{0pt}{3ex} $m_4$ (keV)    & 30     & 7.1  & 40   & 50   & 25000 \\ \hline 
                \rule{0pt}{3ex} $m_{5,6}$ (GeV)& 2.5    & 3.0  & 3.5  & 2.0  & 1.5 \\ \hline 
                \rule{0pt}{3ex} $w$ (GeV)      & 100    & 750  & 250  & 500  & 175 \\\hline
                %            \rule{0pt}{3ex} $\delta_{\rm CP}$ &&&&& \\\hline 
            \end{tabular}
        \end{center} 
    \end{table} 
    
    The various accelerator, beam dump and decay search experiments constrain 
    the combinations 
    \be \label{eq:u2}
    U_e^2 =  \sum_{i=4}^6 |U_{ei}|^2
    \text{~~and~~}
    U_\mu^2 = \sum_{i=4}^6|U_{\mu i}|^2 
    \ee
    of the elements of the active-sterile mixing matrix  $U_{\rm as}$. We can use 
    these sums to investigate the dependence of the neutrino sector of super-weak model 
    on the $R$ matrix, the lightest neutrino mass $m_1$ and the sterile neutrino 
    masses $m_4$, $m_5$ and $m_6$. The sum $U_X^2$ in Eq.~\eqref{eq:u2} represents 
    the weight of sterile components in $\nu_X$ ($X= e$ or $\mu$).
    
    We scanned the parameters of the $R$ matrix over the whole parameter space 
    $(s_{12}, s_{13}, s_{23}) \in [0,1]^3$ to enhance the active-sterile mixings 
    $U_e^2$ and $U_\mu^2$ enough, so that those will be testable at different 
    upcoming experiments. We performed systematic iterative searches by locating 
    the optimal region in the unit cube, followed by a search again in the optimal
    sub-volume with a denser sampling until we reached the desired accuracy of the
    $s_{ij}$ values.
    
    All the benchmark points give valid physics scenarios, and are
    sensitive to different combinations of the experiments.  It turns out
    that the (2,2), (2,3), (3,2) and (3,3) elements dominate $Y_\nu$, as
    they correspond to the heavy right-handed neutrinos $N_2$ and $N_3$. 
    Similarly, the first column in the active-sterile mixing matrix corresponds 
    to mixing of the active neutrinos to $N_1$.  Since $N_1$ is at keV scale,
    active-$N_1$ mixing is stronger than active-$N_2$ and -$N_3$ mixing,
    \be 
    |U_{a 4}| \gg |U_{a 5}|, |U_{a 6}|,\quad a = e,\mu,\tau
    \,.
    \ee 
    
\begin{figure}[t!]
        \centering
        \includegraphics[width=0.43\linewidth]{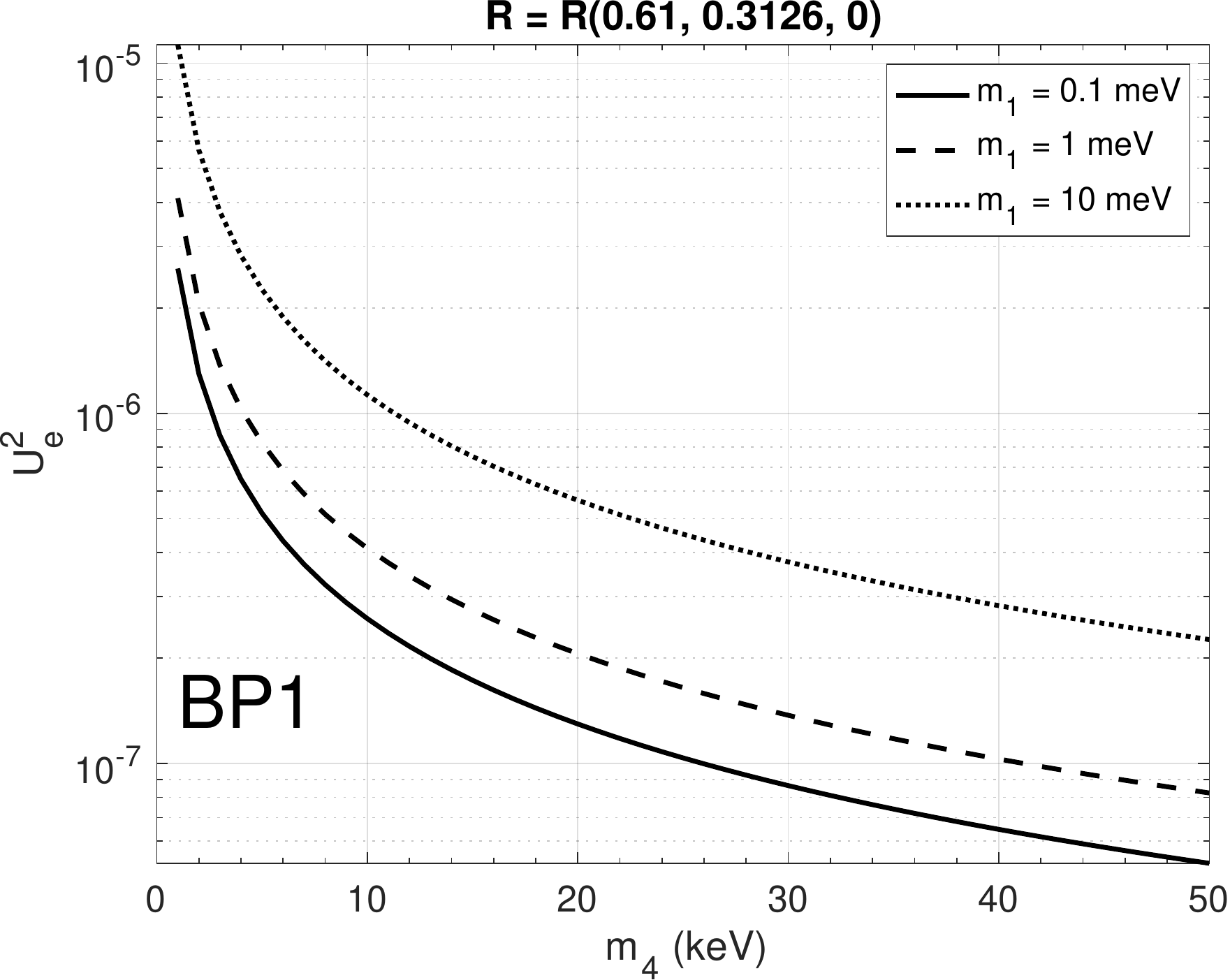}
        \includegraphics[width=0.43\linewidth]{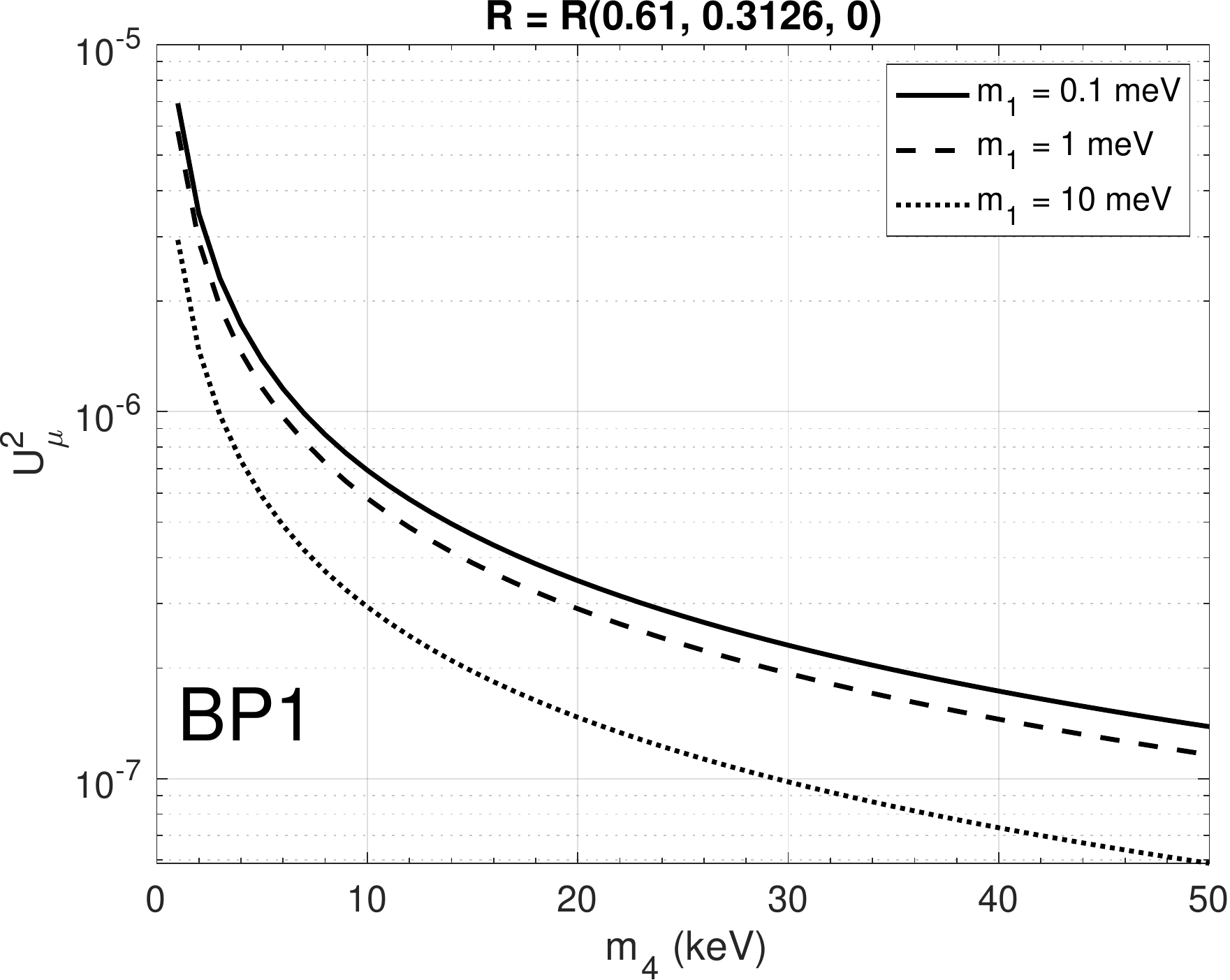}
        \includegraphics[width=0.43\linewidth]{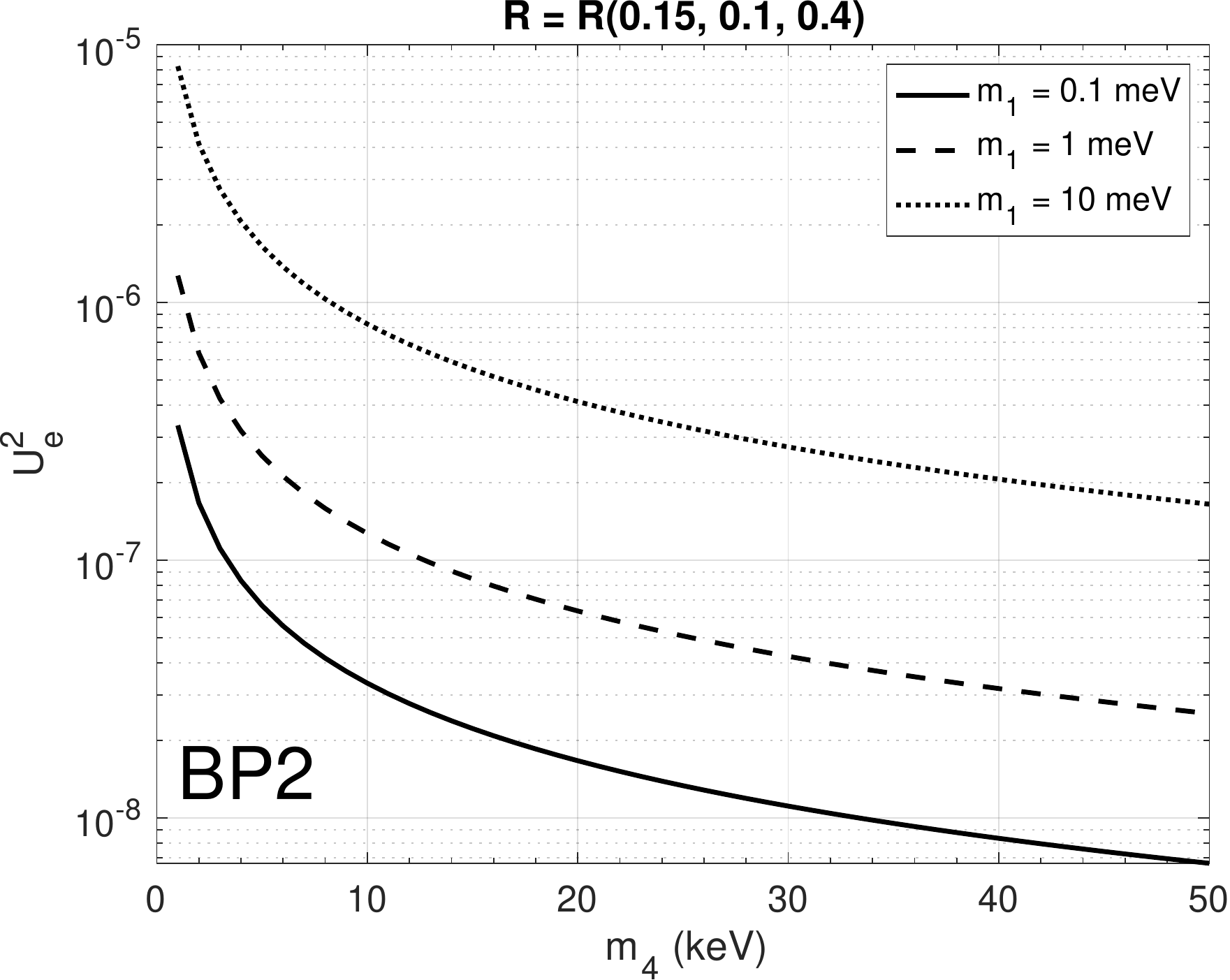}
        \includegraphics[width=0.43\linewidth]{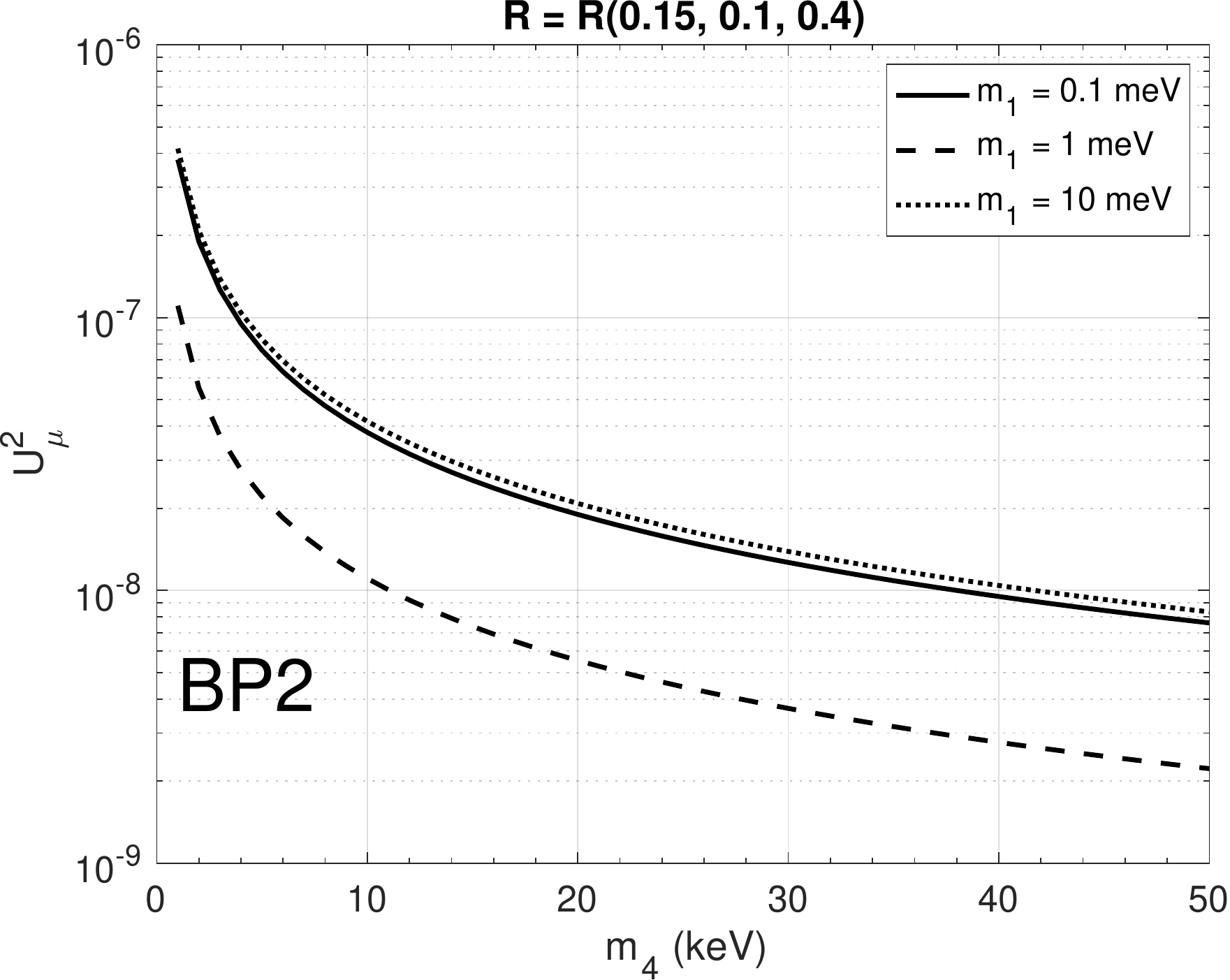}
        \includegraphics[width=0.43\linewidth]{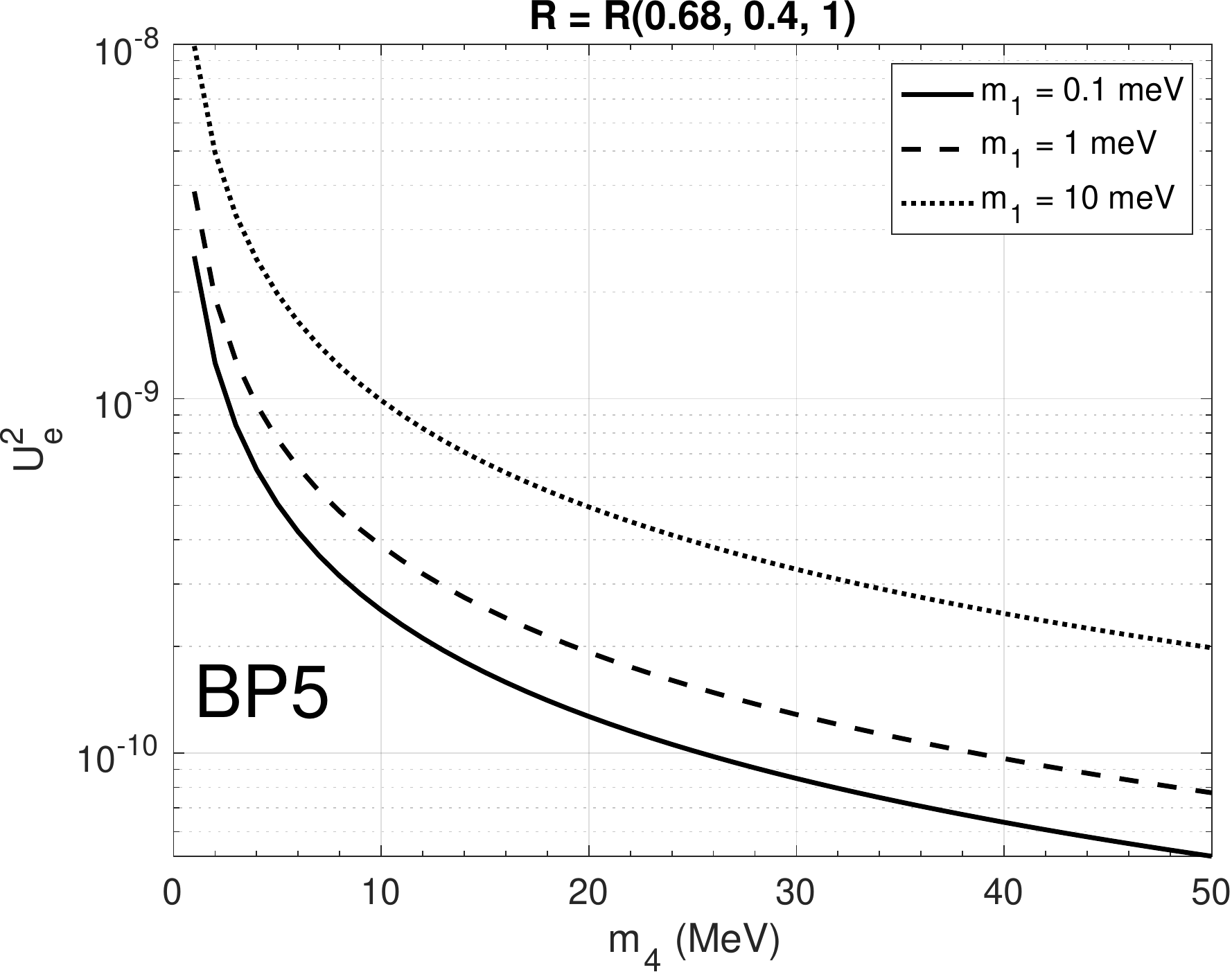}
        \includegraphics[width=0.43\linewidth]{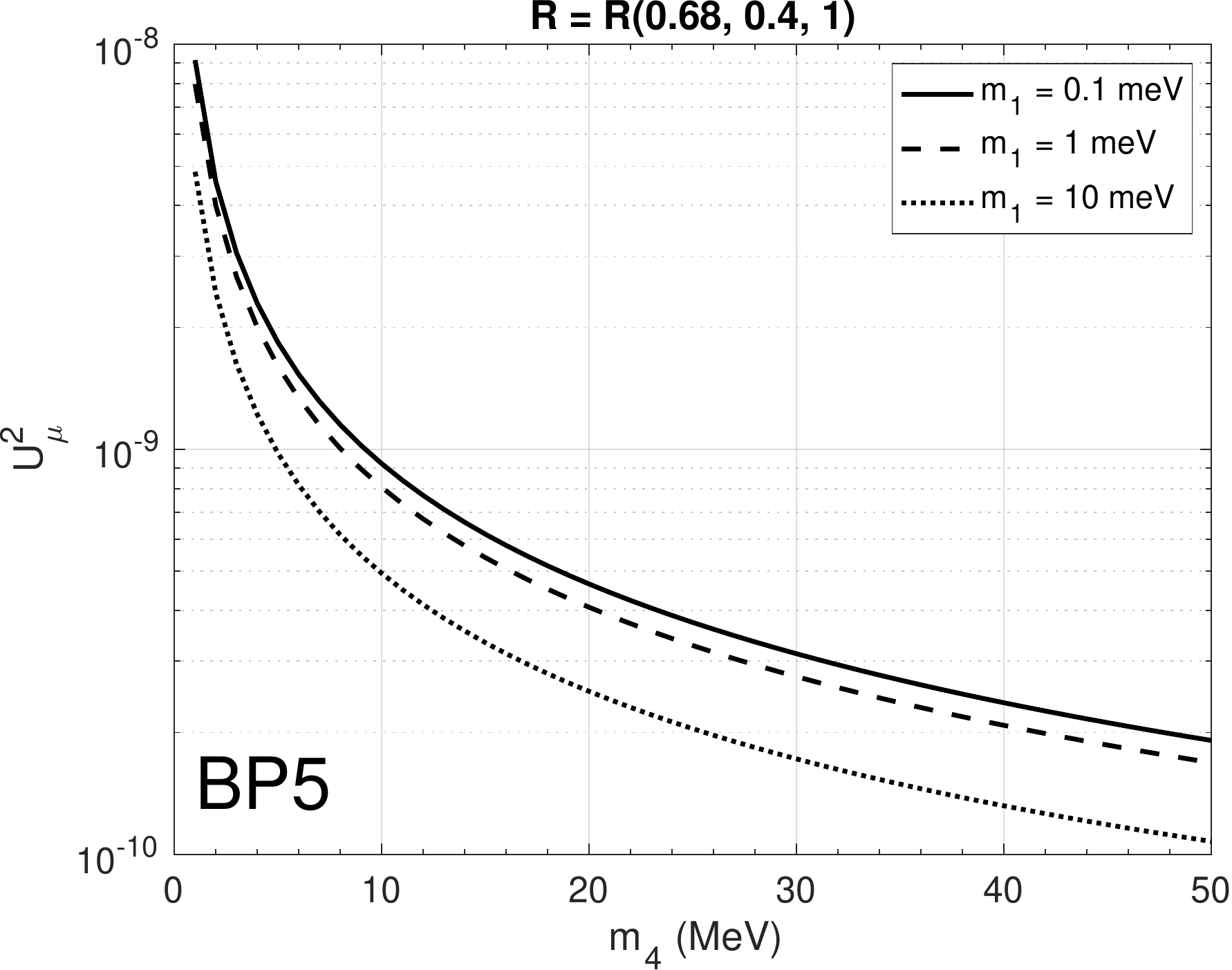}
        \caption{\label{fig:mix}
            Left plots: effective mixing squared of electron neutrino $\nu_e$ to sterile 
            flavours as a function of $m_4$ at benchmark points \textbf{BP1}, 
            \textbf{BP2} and \textbf{BP5}. Right plots: same for $\nu_\mu$.
            We have used $m_1 = 0.1$, 1 and 10\,meV and normal mass hierarchy.}
    \end{figure}

    The values of $U_e^2$ and $U_\mu^2$ at our benchmark points are presented in
    Table~\ref{Out-table} where we have also demonstrated that the relative 
    one-loop corrections to the neutrino masses are at most at the per mill level. 
    In addition, we calculated the leading order correction to the mass $\Delta m_4$
    of the lightest sterile neutrino using Eq.~\eqref{eq:m4-corr}. The correction was 
    found to be less than 1\,\%, with \textbf{BP2} being the exception where the correction is 18.5 \%. The lightest sterile neutrino dominates the effective 
    mixing as it is assumed to be lighter than the other sterile neutrinos by more 
    than four orders of magnitude. Thus, there is no significant dependence of
    $U_e^2$ and $U_\mu^2$ on $m_{5,6}$. In Fig.~\ref{fig:mix} we present the 
    expected effective mixings $U_e^2$ and $U_\mu^2$ as a function of the 
    smallest sterile neutrino mass $m_4$ at selected values of the smallest 
    active neutrino mass $m_1$ for points \textbf{BP1}, \textbf{BP2} and 
    \textbf{BP5}. We know from Eq.~\eqref{eq:Uas-3} that the effective 
    mixing scales as $m_\nu^\text{diag}m_{\rm R}^{-1}$. Departure from the
    simplest choice $R=I_3$ (i.e.~$s_{12} = s_{13} = s_{23} = 0$) can 
    enhance the effective mixing to a degree where it becomes large enough 
    to be accessible in the future experiments.  
    \begin{table}[t!]
        \caption{\label{Out-table}Output values of benchmark points:
            effective active-sterile mixing and neutrino mass correction. For the
            one-loop correction we used $|\tZ| < 10^{-4}$, $\MZp = 20$\,MeV,
            $\Ms = 250$\,GeV and $\sin \theta_S = 0.1$. Note that for \textbf{BP3}
            we have chosen $m_1^\text{tree} = 0$, hence $\Delta m_1^\text{1-loop}/m_1^\text{tree}$ is undefined.}
        \begin{center}
            \begin{tabular}{|c|c|c|c|c|c|}\hline 
                \textbf{Benchmark point }& \textbf{BP1} & \textbf{BP2} & \textbf{BP3} & \textbf{BP4} & \textbf{BP5}\\ \hline 
                %\rule{0pt}{3ex} $U^2$ & $5.0 \times 10^{-7}$ & $1.4 \times 10^{-7}$ & $3.3 \times 10^{-8}$ & $6.3 \times 10^{-8}$ & $7.1 \times 10^{-9}$  \\\hline 
                \rule{0pt}{3ex} $U_e^2$   & $4.0 \times 10^{-7}$ & $1.9 \times 10^{-7}$ & $3.3 \times 10^{-8}$ & $5.0 \times 10^{-8}$ & $3.0 \times 10^{-10}$\\\hline 
                \rule{0pt}{3ex} $U_\mu^2$ & $8.5 \times 10^{-8}$ & $1.2 \times 10^{-8}$ & $5.6 \times 10^{-7}$ & $2.7 \times 10^{-7}$ & $2.4 \times 10^{-10}$\\\hline
                \rule{0pt}{3ex} $\Delta m_1^\text{1-loop}$ (eV) & $1.2 \times 10^{-5}$ & $9.1 \times 10^{-8}$ & 0                    & $1.5 \times 10^{-7}$ & $8.0 \times 10^{-6}$\\\hline 
                \rule{0pt}{3ex} $\Delta m_2^\text{1-loop}$ (eV) & $2.4 \times 10^{-5}$ & $2.4 \times 10^{-5}$ & $2.4 \times 10^{-5}$ & $1.7 \times 10^{-5}$ & $1.7 \times 10^{-5}$ \\\hline 
                \rule{0pt}{3ex} $\Delta m_3^\text{1-loop}$ (eV) & $1.3 \times 10^{-4}$ & $1.4 \times 10^{-4}$ & $3.9 \times 10^{-5}$ & $8.8 \times 10^{-5}$ & $1.1 \times 10^{-4}$ \\\hline 
                \rule{0pt}{3ex} $\Delta m_1^\text{1-loop}/m_1^\text{tree}$ & $1.2 \times 10^{-3}$ & $9.1 \times 10^{-5}$ & undefined            & $1.5 \times 10^{-3}$ & $1.6 \times 10^{-3}$\\\hline 
                \rule{0pt}{3ex} $\Delta m_2^\text{1-loop}/m_2^\text{tree}$ & $1.8 \times 10^{-3}$ & $2.8 \times 10^{-3}$ & $2.8 \times 10^{-3}$ & $2.0 \times 10^{-3}$ & $1.7 \times 10^{-3}$\\\hline 
                \rule{0pt}{3ex} $\Delta m_3^\text{1-loop}/m_3^\text{tree}$ & $2.5 \times 10^{-3}$ & $2.8 \times 10^{-3}$ & $7.6 \times 10^{-4}$ & $1.7 \times 10^{-3}$ & $2.2 \times 10^{-3}$\\\hline 
                \rule{0pt}{3ex} $|\Delta m_4|$ (eV) & 0.01 & 1311 & 212 & 53.0 & 0.008 \\\hline
            \end{tabular}
        \end{center}
    \end{table}
    
\begin{figure}[!t]
        \begin{center}
            \includegraphics[width=0.7\linewidth]{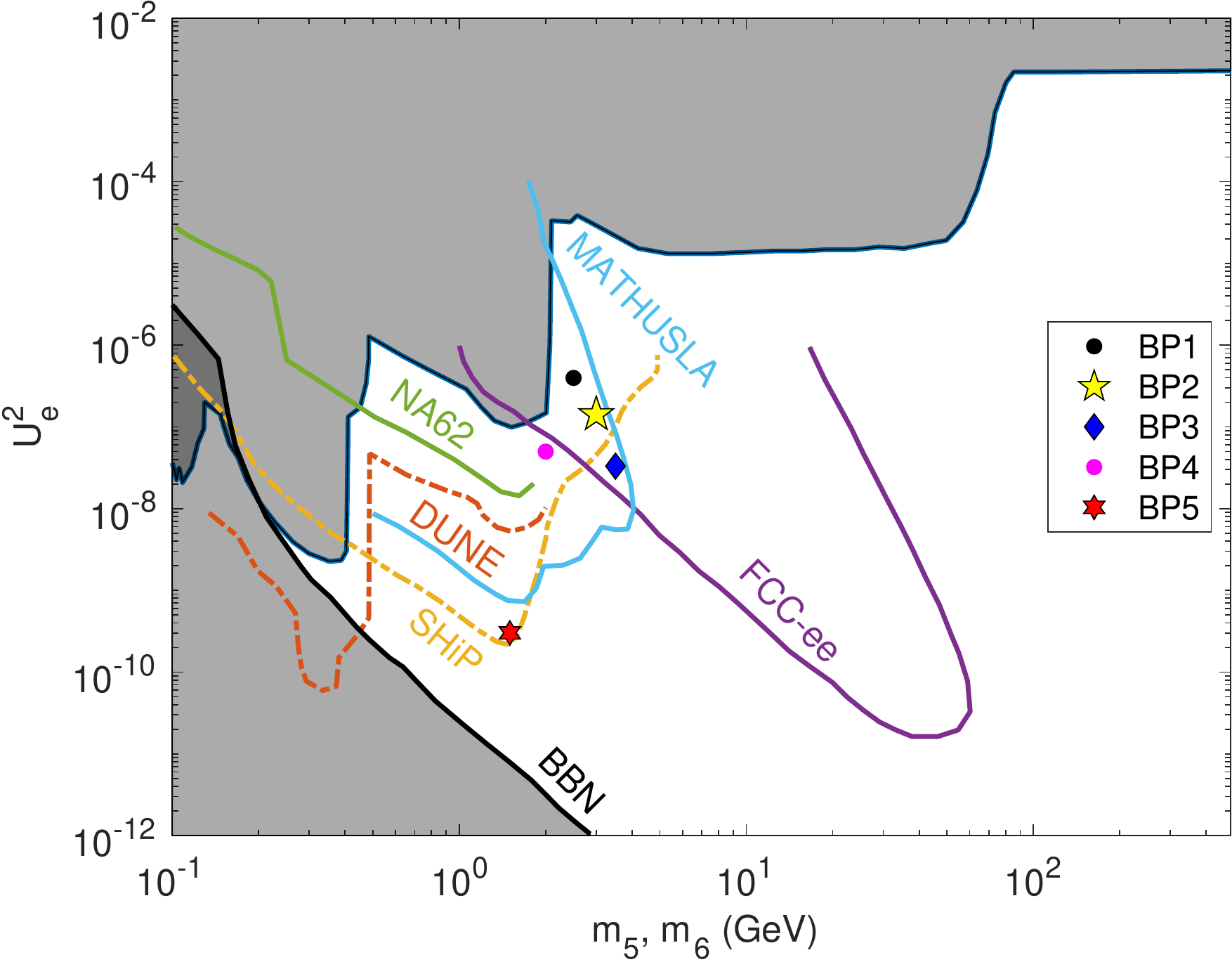}
            \caption{\label{fig-mix2}Constraints in logarithmic $(U_e^2,m_i)$ plane
                $i = 5,6$ from above are given by several experiments (shaded area),
                collected from \cite{Alekhin:2015byh,Drewes:2018gkc,Curtin:2018mvb}. Experimental sensitivities of
                future experiments are given by colored lines.}
        \end{center}
    \end{figure}

\begin{figure}[!h]
        \begin{center}
            \includegraphics[width=0.7\linewidth]{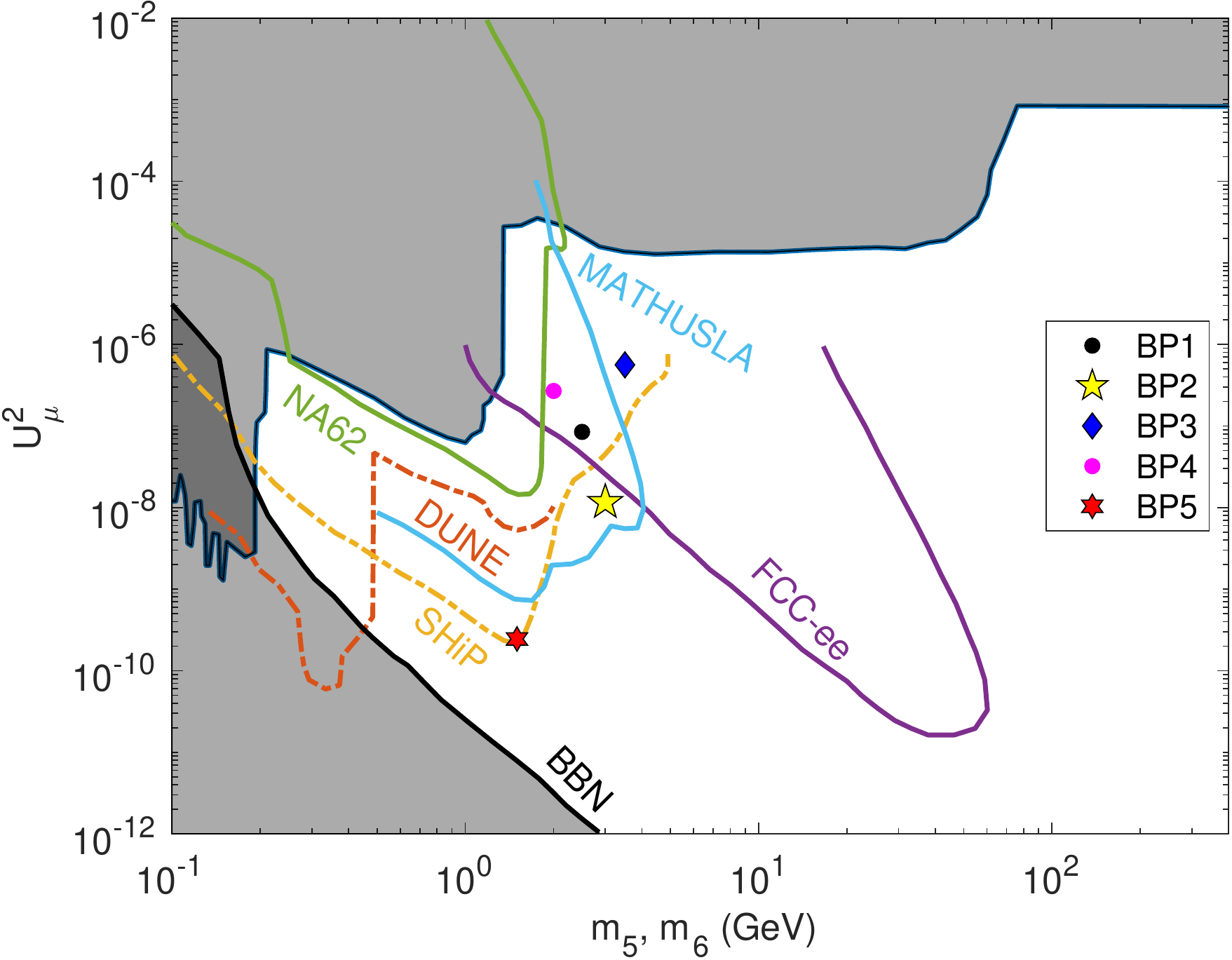}
            \caption{\label{fig-mixMu}As in Fig. \ref{fig-mix2}, but for $U_\mu^2$.}
        \end{center}
    \end{figure}

    We present the current experimental bounds and sensitivities of various 
    future experiments for sterile neutrinos over the $(m_{5,6},U_e^2)$ and
    $(m_{5,6},U_\mu^2)$ planes in Figs.~\ref{fig-mix2} and \ref{fig-mixMu}. 
    The BBN bound in the lower left corner corresponds to the maximal mean lifetime of the sterile neutrino less
    than 1\,s, otherwise it would disrupt the big bang nucleosynthesis
    (BBN) in the early universe \cite{Dolgov:2000jw}. The DUNE curve
    (orange dashed) gives the expected 5-year sensitivity of the DUNE near
    detector with $5\times 10^{21}$ protons on
    target \cite{Adams:2013qkq}. The SHiP line (lemon dashed) represents
    the 90 \% C.L.\ discovery potential of the SHiP experiment
    \cite{Anelli:2015pba}. The FCC-ee exclusion curve (purple solid)
    assumes $10^{12}$ Z boson decays \cite{Blondel:2014bra}. The NA62 bound
    (green solid) is from Ref.~\cite{Drewes:2018gkc} and the MATHUSLA bound
    (blue solid) from Ref.~\cite{Curtin:2018mvb}.  We chose the benchmark
    points in such a way that they all evade the present experimental
    bounds, but can be tested at future experiments (see the legend in
    Figs.~\ref{fig-mix2} and \ref{fig-mixMu}).  We have also checked that 
    active-$N_1$ mixing satisfied the $\beta$ decay electron energy spectrum 
    kink bounds given in \cite{Atre:2009rg}.
    
    \section{Conclusions\label{Sec7}}
    
    In this paper we studied the experimental feasibility of the neutrino sector 
    of the super-weak U(1) extension of the standard model. We demonstrated 
    that there exist benchmark points that are consistent with current observational 
    constraints from neutrino oscillation and scattering experiments.
    At these points the lightest sterile neutrino, with mass of $\rO(10)$\,MeV is
    a viable freeze-out dark matter candidate, while with mass of $\rO(10)$\,keV,
    is so in the freeze-in scenario and can also provide a possible explanation of
    the 3.5\,keV X-ray line observed in galaxy spectra. The nearly degenerate
    heavy sterile neutrinos with mass of $\rO(1)$\,GeV may be probed 
    independently by the MATHUSLA \cite{Curtin:2018mvb} and SHiP
    \cite{Anelli:2015pba,Alekhin:2015byh} experiments in the future. 
    We have shown that the one-loop corrections and seesaw expansion terms provide
    negligible contributions to neutrino masses with our choices of benchmark
    values. 
    
    The extra neutral light $Z'$ gauge boson in the model acts as a mediator 
    between the neutrinos and other fermions, similar to $Z^0$ boson in the SM.
    This $Z'$ boson generates flavour-universal nonstandard interactions, and 
    their strength is consistent with current experimental limits if the gauge 
    sector couplings $g_y'$, $g_z'$ and the mixing angle $|\tZ|$ are less than 
    $\rO(10^{-4})$. Larger values are possible only within a very narrow range 
    if $\tZ > 0$, but disfavoured if the model is to explain the origin of dark 
    matter abundance. The NSI can be made manifest by observing deviations from 
    the neutrino-nucleus or neutrino-electron scattering cross sections as 
    predicted in the standard model.
    
    We have deliberately chosen the masses of the sterile neutrinos $N_2$ and $N_3$ 
    in our benchmark points such that they can fit in the $\nu$MSM scenario 
    \cite{Asaka:2005pn,Shaposhnikov:2006xi,Canetti:2012kh} with the exception of CP
    violation, which is absent in our benchmarks. In addition, we have demonstrated 
    the compatibility of the super-weak extension with the near-future experimental
    sensitivity and its potential solution to the dark matter problem. Hence our 
    conclusions motivate a detailed scan of the allowed parameter space, including non-vanishing CP phases, which we leave for later studies.
    
    \subsection*{Acknowledgments}
    
    We are grateful to Sho Iwamoto for fruitful discussions and to Josu 
    Hernández-García for careful reading of the manuscript. This work was 
    supported by grant K 125105 of the National Research, Development and Innovation 
    Fund in Hungary.
    
    \clearpage
    
    \appendix
    
    \section{Flavour-universal NSI and non-unitary mixing}
    \label{app:A}
    
    It is possible to perform a phase rotation in effective neutrino oscillation Hamiltonian, 
    but it turns out that we cannot rotate away the NSI fully even if the NSI matrix
    \be 
    \varepsilon_{\rm true} = \begin{pmatrix}
        \varepsilon^{\rm m}_{ee} & \varepsilon^{\rm m}_{e\mu} & \varepsilon^{\rm m}_{e\tau} \\
        \varepsilon^{\rm m}_{\mu e} & \varepsilon^{\rm m}_{\mu\mu} & \varepsilon^{\rm m}_{\mu \tau} \\
        \varepsilon^{\rm m}_{\tau e} & \varepsilon^m_{\tau \mu} & \varepsilon^{\rm m}_{\tau\tau}
    \end{pmatrix}
    \ee 
    is isotropic (that is, proportional to unit matrix), since the active neutrino mixing matrix, 
    which rotates the neutrino flavour basis to neutrino mass basis, is non-unitary. Instead,
    the NSI is merely suppressed by unitarity deviations. The transformed NSI matrix is actually
    \be 
    \varepsilon'= N^\dagger\varepsilon_\text{true}N,
    \ee 
    where $N$ is 3 $\times$ 3 non-unitary matrix in neutrino flavour basis. In the super-weak model, 
    the new NSI matrix is not isotropic, but the isotropic part can be isolated and removed. 
    One gets an interesting interplay from the NSI and non-unitarity, which means in practise 
    that in long baseline neutrino oscillations (LBNO) the effect from these two sources is 
    indistinguishable. Consider an example: suppose we measure a 2 \% distortion in neutrino 
    oscillation probabilities, and we cannot fit it in standard three-neutrino no-BSM framework. 
    This distortion may be caused fully by the NSI, fully by non-unitarity or by both in unknown 
    ratio. In this case the LBNO experiments constrain only the combined effect. Short-baseline 
    experiments in turn constrain non-unitarity.
    
    Let $N \equiv (I-\alpha)U_\text{PMNS}$ be the non-unitary neutrino mixing matrix. The deviation matrix
    \be 
    \alpha = \begin{pmatrix}
        \alpha_{ee} & 0 & 0 \\ \alpha_{\mu e} & \alpha_{\mu\mu} & 0 \\ \alpha_{\tau e} & \alpha_{\tau \mu} & \alpha_{\tau\tau} 
    \end{pmatrix}
    \ee 
    has small elements, so we may ignore $\rO(\alpha^2)$ part. Thus,
    \be
    H_\text{NSI} = N^\dagger \begin{pmatrix}
        \varepsilon & 0 & 0 \\ 0 & \varepsilon & 0 \\ 0 & 0 & \varepsilon
    \end{pmatrix} N \approx \varepsilon I - U_\text{PMNS}^\dagger(\alpha + \alpha^\dagger)U_\text{PMNS}\varepsilon
    \,.
    \ee
    The first term is an irrelevant isotropic phase term in neutrino oscillations. 
    The second term is neither isotropic nor diagonal. It is the leading order part of the NSI,
    which cannot be cancelled by phase-rotation, suppressing the NSI by about
    $|\alpha| \sim |U_{\rm a s}| \lesssim 10^{-3}$. Thus one must consider two different 
    sets of bounds for oscillation and scattering experiments: the suppressed bounds for neutrino 
    oscillations and unsuppressed ones for scattering.
    
    \section{Light NSI mediator}
    
    Upon deriving the NSI operator, we integrate out the mediator by assuming that the propagator 
    may be approximated by negative inverse mass squared:
    $$
    \frac{1}{p^2-\MZp^2} \approx \frac{-1}{\MZp^2}.
    $$
    This is valid only if the momentum transfer in the scattering is small compared to the 
    mass of the mediator, $\MZp$. In neutrino propagation in matter only forward scattering 
    interactions are relevant. In those cases the momentum transfer via mediator is zero. 
    With nonzero momentum transfer the outgoing neutrino will be deflected away from the 
    source-to-detector trajectory even if the momentum transfer is tiny. It is therefore 
    possible to consider NSI with light mediators. However for scattering experiments one 
    needs to have mediator mass at least about 10\,MeV in order to have a meaningful 
    approximation \cite{Esteban:2018ppq,Farzan:2015hkd}.
    
    \bibliography{sweak}
\end{document}